\documentclass[a4paper,10pt]{article}
\usepackage{indentfirst}
\usepackage[colorlinks,linkcolor=black,anchorcolor=black,citecolor=blue,urlcolor=blue]{hyperref}
\usepackage{graphics}
\usepackage{epsfig}
\usepackage{epstopdf}
\usepackage[utf8]{inputenc}
\usepackage{url}     
\usepackage{slashed} 
\usepackage{booktabs}
\usepackage{mathrsfs}
\usepackage{epsfig}
\usepackage{dcolumn}
\usepackage{bm}
\usepackage{amsmath}
\usepackage{multirow}
\usepackage{amsmath}		
\usepackage{amssymb}		
\usepackage{caption, subcaption}
\usepackage{comment,color}
\usepackage{cite}

\begin{document}

\thispagestyle{empty} \vspace*{0.8cm}\hbox
to\textwidth{\vbox{\hfill\huge\sf \hfill}}
\par\noindent\rule[3mm]{\textwidth}{0.2pt}\hspace*{-\textwidth}\noindent
\rule[2.5mm]{\textwidth}{0.2pt}


\begin{center}
\LARGE\bf The mass-degenerate SM-like Higgs and anomaly of $(g-2)_\mu$ in $\mu$-term extended NMSSM
\end{center}

\begin{center}
Liangliang Shang, XiaoFeng Zhang, and Zhaoxia Heng\footnote{zxheng@htu.edu.cn}
\end{center}

\begin{center}
\begin{footnotesize} \sl
{School of Physics, Henan Normal University, Xinxiang 453007, China} \\
\end{footnotesize}
\end{center}

\vspace*{2mm}

\begin{center}
\begin{minipage}{15.5cm}
\parindent 20pt\footnotesize
\begin{abstract}
We chose the $\mu$-term extended next-to-minimal supersymmetric standard model ($\mu$NMSSM)  for this work, 
and we perform a phenomenological study based on the assumption that the
	observed Standard Model (SM)-like Higgs is explained by the presence of a double overlapping
	resonance and in light of the recent $(g-2)_\mu$ result.
The study also takes into account a variety of experimental results, including Dark Matter (DM) direct detections  and results from sparticle searches at the Large Hadron Collider (LHC). 
We study the properties of DM confronted with the limits from DM direct detections.
As a second step, we focus our attention on
the properties of the mass-degenerate SM-like Higgs bosons and on explaining the anomaly of $(g-2)_\mu$. 
We conclude that the anomaly of $(g-2)_\mu$ can be explained in the scenario with two mass-degenerate SM-like Higgs, and
there are samples that meet the current constraints and fit $1-\sigma$ anomalies of Higgs data.

\end{abstract}
\end{minipage}
\end{center}

\begin{center}
\begin{minipage}{15.5cm}
\begin{minipage}[t]{2.3cm}{\bf Keywords:}\end{minipage}
\begin{minipage}[t]{13.1cm}
$\mu$NMSSM; mass-degenerate SM-like Higgs; anomaly of ($g-2)_\mu$; DM
\end{minipage}\par\vglue8pt
\end{minipage}
\end{center}

\newpage
\tableofcontents
\section{Introduction}
\label{intro}
The publication of the latest data by the Fermi National Accelerator Laboratory (FNAL) and Brookhaven National Laboratory (BNL) show a 4.2$\sigma$ discrepancy from the Standard Model (SM) prediction \cite{1,2,Aoyama:2020ynm,c3,c4,c5,c6,c7,c8,c9,c10,c11,c12,c13,c14,c15,c16,c17,c18,c19,c20,c21,c22,c23,c24,c25,c26,c27,c28,c29,c30,c31,c32,c33,c34,c35,c36,c37}.
When the experimental accuracy is further improved, the  measurement of the muon anomalous magnetic moment $(g - 2)_\mu$  
is likely to become a breakthrough of new physics\cite{3a0,3a1,3a2,3a3}.
So the interpretation of $(g - 2)_\mu$ anomaly has also become an important task of new physics. For examples, low-energy supersymmetry (SUSY) can naturally explain the anomaly of  $(g - 2)_\mu$\cite{li1,
		li4,li5,li6,li7,li8,li9,li10,
	li11,li12,li13,li14,li15,
	li17,li18,li19,li20,
	li21,li22,li23,li24,li25,li26,li27,li28,li29,li30,
	li31,li32,li33,li34,li35,li36,li37,li38,li39,li40,
	li41,li42,li43,li44,li45,li46,li47,li48,li49,li50,
	li51,li52,li53,li54,li55,li56,li57,li58,li59,li60,
	li61,li62,li63,li64,li65,li66,li67,li68,li69,li70,
	li71,li72,li73,li74,li75,li76,li77,li78,li79,li80,
	li81,li82,li83,li84,li85,li86,li87,li88,li89,li90,
	Cao:2022chy,Chakraborti:2022sbj}. 

SUSY models have been widely studied due to its multiple theoretical advantages,  such as the explanation of the gauge
hierarchy problem, the unification of gauge coupling,  and natural solutions to the Dark Matter (DM) mystery.  As the most economical realization of SUSY,  
the minimal supersymmetric standard model (MSSM)\cite{3a3b2} with R-parity conservation can provide a viable candidate of DM, which is lightest neutralino as the lightest
 supersymmetric particle (LSP). However, considering the constraints from DM relic density and DM direct detections, MSSM is strongly restricted, 
 and there are also some problems in the MSSM, such as the $\mu$-problem  and little hierarchy problem, which can be solved in the 
 next-to-minimal supersymmetric standard model (NMSSM)\cite{3a3b3,3a3b4}.  NMSSM extends the MSSM by one singlet Higgs superfield $\hat S$, 
 which develops a vacuum expectation value (VEV) to generate an effective $\mu$-term.  Furthermore, because of  the enlarged Higgs sector, the SM-like Higgs boson mass can be easily interpreted\cite{3a3b5}. In the NMSSM with a $Z_3$ symmetry ($Z_3$-NMSSM), to realize a singlino-dominant DM,  $2|\kappa|/\lambda$ must be
  less than 1\cite{42}. However, the situation is different in the general NMSSM (GNMSSM). We consider a simplified version of GNMSSM, i.e., the $\mu$-term extended $Z_3$-NMSSM ($\mu$NMSSM)\cite{17}. In the $\mu$NMSSM, the mass ratio of singlino and higgsino is no longer equal to $2|\kappa|/\lambda$,  so the values of
   $\kappa$ have a wider range compared to the case in the $Z_3$-NMSSM\cite{16}. The parameter  $\kappa$ has an important effect on the singlet fields’ self-interactions,
which can be  entirely responsible for the DM density. Therefore, considering the latest DM experimental constraints, the $\mu$NMSSM has a wider parameter space than the $Z_3$-NMSSM.

In July 2012, both the ATLAS and CMS collaborations announced a scalar with mass near 125 GeV\cite{29,30}.  Combined measurements of Higgs boson production cross sections and decay branching fractions show no significant deviations from Standard Model (SM) predictions. However, for the Higgs production process  in association with a top quark pair followed by the decay mode to vector boson ($H\longrightarrow WW$) or photon ($H\longrightarrow \gamma\gamma$)\cite{3,4,5,6,7}, there is a discrepancy between the SM prediction and the experimental data. Many theories attempt to interpret the observed data at the LHC.   In the $\mu$NMSSM, either the lightest CP-even Higgs boson $H_1$ or the next-to-lightest CP-even Higgs boson $H_2$ can be SM-like with mass near 125 GeV.  In fact,  it is also possible to have $H_1$ and  $H_2$ nearly degenerate with mass near 125 GeV\cite{3a3b6,3a3b7,8,34, Shang:2020uog}, so that the observed signal at the LHC is actually a superposition of
two individual peaks, and these two peaks cannot be independently resolved. In our work we focus on the scenario with two mass-degenerate Higgs boson with mass near 125 GeV in the $\mu$NMSSM considering various constraints including the DM relic density and DM direct detection limits. Furthermore, we  try to explain the anomaly of  $(g - 2)_\mu$ in this scenario.

The paper is structured as follows. In Section 2, we briefly introduce the basic properties of $\mu$NMSSM including the Higgs, neutralino and chargino sectors. Then we study the observables of the 125 GeV Higgs boson at the LHC, DM and  $(g - 2)_\mu$ in the $\mu$NMSSM. We also describe our scanning strategy. In Section 3, we investigate the properties of DM confronted with DM direct detection limits. In Section 4, we show the numerical results in the scenario with two mass-degenerate Higgs boson and explain the anomaly of $(g-2)_\mu$. Finally, we give a summary in Section 5.

\section{\label{sec:theo}Theoretical preliminaries}

\subsection{Basics of $\mu$NMSSM}

To address a number of flaws in MSSM, such as the $\mu$ problem, we consider the next-to-minimal extension. 
Compared with MSSM, NMSSM introduces a singlet Higgs field $\widehat{S}$. The superpotential of the general NMSSM (GNMSSM) is given by
\begin{equation}
	W_{\text {GNMSSM}}=W_{\text {Yukawa }}+(\mu+\lambda \widehat{S}) \widehat{H}_{u} \cdot \widehat{H}_{d}+\xi_{F} \widehat{S}+\frac{1}{2} \mu^{\prime} \widehat{S}^{2}+\frac{\kappa}{3} \widehat{S}^{3}
\end{equation}
where the Yukawa terms $W_{\text {Yukawa }}$ are as follows \cite{3a3b4},
\begin{equation}
	W_{\text {Yukawa }}=h_{u} \widehat{Q} \cdot \widehat{H}_{u} \widehat{U}_{R}^{c}+h_{d} \widehat{H}_{d} \cdot \widehat{Q} \widehat{D}_{R}^{c}+h_{e} \widehat{H}_{d} \cdot \widehat{L} \widehat{E}_{R}^{c}
\end{equation}
Among them, $\lambda$ and $\kappa$ are dimensionless couplings, $\mu$ and $\mu^{\prime}$ are supersymmetric mass terms, and $\xi_{F}$ is the supersymmetric tadpole term of mass square dimension. 

In the above formula, the $\mu$, $\mu^{\prime}$ and $\xi_{F}$ terms break the $\mathbb{Z}_3$-symmetry, and the $\frac{\kappa}{3}$ term conserves the $\mathbb{Z}_3$-symmetry while breaking the PQ-symmetry. 
Some articles use the terms $\mu$ and $\xi_{F}$ to explain the tadpole problem \cite{20} and the cosmological domain-wall problem\cite{21,22,23}. 
To maintain the purpose of our research without sacrificing generality, we set $\mu^{\prime}$ = $\xi_{F}$ = 0. This is known as the $\mu$-term extended NMSSM ($\mu$NMSSM).
The extra $\mu$-term is related to the non-minimal supergravity coupling $\chi$ via the gravitino mass as $\mu = \frac{3}{2} m_{3/2} \chi $, where $m_{3/2}$ denotes the gravitino mass\cite{24}.
The superpotential and soft supersymmetry-breaking terms of $\mu$NMSSM are given below\cite{25,26},
\begin{equation}
	\begin{aligned}
		W_{\mu \mathrm{NMSSM}} &=W_{\text {Yukawa }}+(\lambda \hat{S}+\mu) \hat{H}_{u} \cdot \hat{H}_{d}+\frac{1}{3} \kappa \hat{S}^{3} ,\\
		\qquad -\mathcal{L}_{\mathrm{soft}} &=\left[A_{\lambda} \lambda S H_{u} \cdot H_{d}+\frac{1}{3} A_{\kappa} \kappa S^{3}+B_{\mu} \mu H_{u} \cdot H_{d}+h . c .\right] \\
		&+m_{H_{u}}^{2}\left|H_{u}\right|^{2}+m_{H_{d}}^{2}\left|H_{d}\right|^{2}+m_{s}^{2}|S|^{2}+\cdots.
	\end{aligned}
\end{equation}

The breaking of electroweak symmetry allows Higgs field to obtain non-zero vacuum expected values (vevs) $<H_{u}^0>$ = $v_u/\sqrt{2}$, $<H_d^0>$ = $v_d/\sqrt{2}$, $<H_s^0>$ = $v_s/\sqrt{2}$. The expression of the Higgs fields can be written as
\begin{equation}
	\begin{gathered}
		H_{u}=\left(\begin{array}{c}
			h_{u}^{+} \\
			h_{u}
		\end{array}\right)=\left(\begin{array}{c}
			\eta_{u}^{+} \\
			\frac{1}{\sqrt{2}}\left(v_{u}+\sigma_{u}+i \phi_{u}\right)
		\end{array}\right), \quad H_{d}=\left(\begin{array}{c}
			h_{d} \\
			h_{d}^{-}
		\end{array}\right)=\left(\begin{array}{c}
			\frac{1}{\sqrt{2}}\left(v_{d}+\sigma_{d}+i \phi_{d}\right) \\
			\eta_{d}^{-}
		\end{array}\right) ,\\
		S=\frac{1}{\sqrt{2}}\left(v_{s}+\sigma_{s}+i \phi_{s}\right).
	\end{gathered}
\end{equation}
The ratio of the two Higgs doublet vevs defines the parameter tan$\beta$ = $v_u$ / $v_d$ with $v=\sqrt{v_{u}^{2}+v_{d}^{2}}$ = 246 GeV, and the effective $\mu$-term $\mu_{eff}$  is generated by $\mu_{eff}$ = $\lambda v_s/\sqrt{2}$.

We set $B_\mu$ = 0 in the soft breaking term since it plays a minor role in our work,  then we get a Higgs potential as follows,
\begin{equation}
	\begin{aligned}
		V_{\text {Higgs }}=&\left(m_{H_{d}}^{2}+\left(\mu+\lambda S\right)^{2}\right)\left|H_{d}\right|^{2}+\left(m_{H_{u}}^{2}+\left(\mu+\lambda S\right)^{2}\right)\left|H_{u}\right|^{2} \\
		&+\left(\kappa S^{2}+\lambda H_{u} \cdot H_{d}\right)^{2}+\frac{g_{2}^{2}}{2}\left|H_{d}^{\dagger} H_{u}\right|^{2}+\frac{g_{1}^{2}+g_{2}^{2}}{8}\left(\left|H_{d}\right|^{2}-\left|H_{u}\right|^{2}\right)^{2} \\
		&+m_{S}^{2} S^{2}+2 \lambda A_{\lambda} S H_{u} \cdot H_{d}+\frac{2}{3} \kappa A_{\kappa} S^{3}
	\end{aligned}
\end{equation}

For convenience, we use the field combinations  $H_{\mathrm{SM}} \equiv \sin \beta \operatorname{Re}\left(H_{u}\right)+\cos \beta \operatorname{Re}\left(H_{d}\right)$, $H_{\mathrm{NSM}} \equiv \cos \beta \operatorname{Re}\left(H_{u}\right)-\sin \beta \operatorname{Re}\left(H_{d}\right)$, and $A_{\mathrm{NSM}} \equiv \cos \beta \operatorname{Im}\left(H_{u}\right)-\sin \beta \operatorname{Im}\left(H_{d}\right)$, then the mass matrix of CP-even Higgs bosons in the basis $\left(H_{\mathrm{NSM}}, H_{\mathrm{SM}}, \operatorname{Re}(S)\right)$ can be written as follows,  
\begin{equation}
	\begin{aligned}
		&\mathcal{M}_{S, 11}^{2}=\frac{2 \mu_{\mathrm{eff}}\left(\lambda A_{\lambda}+\kappa \mu_{\mathrm{eff}}\right)}{\lambda \sin 2 \beta}+\frac{1}{2}\left(2 m_{Z}^{2}-\lambda^{2} v^{2}\right) \sin ^{2} 2 \beta \\
		&\mathcal{M}_{S, 12}^{2}=-\frac{1}{4}\left(2 m_{Z}^{2}-\lambda^{2} v^{2}\right) \sin 4 \beta \\
		&\mathcal{M}_{S, 13}^{2}=-\frac{1}{\sqrt{2}}\left(\lambda A_{\lambda}+2 \kappa \mu_{\mathrm{eff}}\right) v \cos 2 \beta \\
		&\mathcal{M}_{S, 22}^{2}=m_{Z}^{2} \cos ^{2} 2 \beta+\frac{1}{2} \lambda^{2} v^{2} \sin ^{2} 2 \beta \\
		&\mathcal{M}_{S, 23}^{2}=\frac{v}{\sqrt{2}}\left(2 \lambda \mu_{\mathrm{eff}}+2 \lambda \mu-\left(\lambda A_{\lambda}+2 \kappa \mu_{\mathrm{eff}}\right) \sin 2 \beta\right), \\
		&\mathcal{M}_{S, 33}^{2}=\frac{\lambda A_{\lambda} \sin 2 \beta}{4 \mu_{\mathrm{eff}}} \lambda v^{2}+\frac{\mu_{\mathrm{eff}}}{\lambda}\left(\kappa A_{\kappa}+\frac{4 \kappa^{2} \mu_{\mathrm{eff}}}{\lambda}\right)-\frac{\lambda \mu}{2 \mu_{\mathrm{eff}}} \lambda v^{2}
	\end{aligned}
\end{equation}
and those for CP-odd Higgs bosons in the basis $\left(A_{\mathrm{NSM}}, \operatorname{Im}(S)\right)$ are as follows
\begin{equation}
	\begin{aligned}
		\mathcal{M}_{P, 11}^{2} &=\frac{2 \mu_{\mathrm{eff}}\left(\lambda A_{\lambda}+\kappa \mu_{\mathrm{eff}}\right)}{\lambda \sin 2 \beta}, \quad \mathcal{M}_{P, 12}^{2}=\frac{v}{\sqrt{2}}\left(\lambda A_{\lambda}-2 \kappa \mu_{\mathrm{eff}}\right) \\
		\mathcal{M}_{P, 22}^{2} &=\frac{\left(\lambda A_{\lambda}+4 \kappa \mu_{\mathrm{eff}}\right) \sin 2 \beta}{4 \mu_{\mathrm{eff}}} \lambda v^{2}-\frac{3 \mu_{\mathrm{eff}}}{\lambda} \kappa A_{\kappa}-\frac{\lambda \mu}{2 \mu_{\mathrm{eff}}} \lambda v^{2}
	\end{aligned}
\end{equation}
By diagonalizing the mass matrix $\mathcal{M}^2_S$ and $\mathcal{M}^2_P$ using the unitary rotations $V$ and $V_P$, we can get the mass eigenstates $H_i(i=1,2,3)$ and $A_i(i=1,2)$. The mass of the charged Higgs bosons $m_{H^{\pm}}$ can be written as follows
\begin{equation}
 m_{H^{\pm}}^{2}=m_{W}^{2}-v^{2} \lambda^{2}+\frac{\mu_{\mathrm{eff}}}{\cos \beta \sin \beta}\left(\frac{\kappa}{\lambda} \mu_{\mathrm{eff}}+A_{\lambda}\right)
\end{equation}

The neutralino sector in the $\mu$NMSSM consists of Bino field $\tilde{B}^{0}$, Wino field $\tilde{W}^{0}$, Higgsino fields $\tilde{H}_{d}^{0}, \tilde{H}_{u}^{0}$ and Singlino field $\tilde{S}^{0}$.  In the basis $\psi^{0}=\left(-i \tilde{B}^{0},-i \tilde{W}^{0}, \tilde{H}_{d}^{0}, \tilde{H}_{u}^{0}, \tilde{S}^{0}\right)$, the mass matrix can be given by \cite{25}
\begin{equation}
	M_{\tilde{\chi}^{0}}=\left(\begin{array}{ccccc}
		M_{1} & 0 & -m_{Z} \sin \theta_{\mathrm{w}} \cos \beta & m_{Z} \sin \theta_{\mathrm{w}} \sin \beta & 0 \\
		\cdot & M_{2} & m_{Z} \cos \theta_{\mathrm{w}} \cos \beta & -m_{Z} \cos \theta_{\mathrm{w}} \sin \beta & 0 \\
		\cdot & \cdot & 0 & -\left(\mu+\mu_{\mathrm{eff}}\right) & -\lambda v \sin \beta \\
		\cdot & \cdot & \cdot & 0 & -\lambda v \cos \beta \\
		\cdot & \cdot & \cdot & \cdot & 2 \frac{\kappa}{\lambda} \mu_{\mathrm{eff}}
	\end{array}\right)
\end{equation}
where $\theta_W$ denotes the Weinberg angle, $M_{1}$ and $M_{2}$ denote the gaugino soft  breaking masses. By a rotation matrix $N$, we can get the mass eigenstate $\tilde{\chi}_i^{0}(i=1-5)$, which are labeled in mass-ascending order. It can clearly see that the higgsino mass in the  $\mu$NMSSM is determined by the parameter $\mu_{tot}=\mu+\mu_{\mathrm{eff}}$, and the singlino mass mainly depends on  $2 \frac{\kappa}{\lambda} \mu_{\mathrm{eff}}$.

In the basis $\psi^{\pm}=\left(\tilde{W}^{+}, \tilde{H}_{u}^{+}, \tilde{W}^{-}, \tilde{H}_{d}^{-}\right)
$, the chargino mass matrix is written as
\begin{equation}
	M_{\tilde{\chi}^{\pm}}=\left(\begin{array}{cc}
		M_{2} & \sqrt{2} m_{W} \sin \beta \\
		\sqrt{2} m_{W} \cos \beta & \mu+\mu_{\mathrm{eff}}
	\end{array}\right)
\end{equation}
By rotation matrix $U^c$ and $V^c$, we can get the chargino mass eigenstate $\chi^{\pm}_i (i=1,2)$.

For convenience, we have selected the following independent parameters as input parameters
\begin{equation}
\lambda, \quad \kappa, \quad A_{\lambda}, \quad A_{\kappa}, \quad \mu, \quad \mu_{\mathrm{tot}}, \quad \tan \beta
\end{equation}

\subsection{Observables of the 125 GeV Higgs bosons}
The ATLAS and CMS collaborations discovered a 5$\sigma$ signal for a Higgs-like resonance with mass around 125 GeV \cite{29,30}. However, these signal channels deviate by 1-2$\sigma$ from the SM prediction. Of course, as the experiment's precision improves, we will be able to confirm the true nature of the discovered Higgs boson. But firstly, we need to see if the two mass-degenerate Higgs hypothesis can provide a better explanation for the current experiment. The enhancement of final state $\gamma\gamma$ under the production modes of gluon fusion ($ggH$) and vector boson fusion ($VBF$) is one of the deviations between the current experimental data and the SM prediction.

We usually focus on the product of the production cross section $\sigma_i$ and the decay branching ratio $B_f$, so we define the observable $O_{i f}$,
\begin{equation}
\label{R}
\begin{split}
	O_{i f} &= \sum_{\alpha} O^{\alpha}_{i f} \\
	O^{\alpha}_{i f} & =\sigma^{H_\alpha}_{i} B^{H_\alpha}_{f}
\end{split}
\end{equation}
where $i$ denotes the production modes: gluon fusion ($ggH$), vector boson fusion ($VBF$), associated production with a Z or W boson (VH), and associated production with a top quark pair (ttH), $f$ denotes the decay modes: $\gamma\gamma$, $WW$, $\tau\tau$, $bb$, etc., $\alpha$ denotes the index of the resonance.
Eq.(\ref{R}) is in general case, with a sum over the index of the resonance, and we can specify it to the case of a
	single resonance.
We list major $O_{i f}$ in Table \ref{rif} and we could arbitrarily expand the table based on the process discovered at the LHC experiment, or we can empty the table that was not discovered at the LHC experiment.
The error size of these data varies greatly due to the number of collider events and other factors. 
\begin{table}[!htb]
	\centering
	\begin{tabular}{l|lllll}
		& $\mathrm{H} \rightarrow \gamma \gamma$  &$\mathrm{H} \rightarrow WW$ & $\mathrm{H} \rightarrow ZZ$ & $\mathrm{H} \rightarrow \tau\tau$ &$\mathrm{H} \rightarrow bb$ \\ \hline
		ggH & $O_{ggH,\gamma\gamma}$ & $O_{ggH,WW}$   & $O_{ggH,ZZ}$ & $O_{ggH,\tau\tau}$&$O_{ggH,bb}$   \\
		VBF & $O_{VBF,\gamma\gamma}$ & $O_{VBF,WW}$   & $O_{VBF,ZZ}$ & $O_{VBF,\tau\tau}$&$O_{VBF,bb}$   \\
		VH  & $O_{VH,\gamma\gamma}$ & $O_{VH,WW}$   & $O_{VH,ZZ}$ & $O_{VH,\tau\tau}$&$O_{VH,bb}$   \\
	ttH	 & $O_{ttH,\gamma\gamma}$ & $O_{ttH,WW}$   & $O_{ttH,ZZ}$ & $O_{ttH,\tau\tau}$&$O_{ttH,bb}$  
	\end{tabular}

\caption{Rows represent the Higgs boson production modes:  gluon fusion ($ggH$), vector boson fusion ($VBF$), associated production with a Z or W boson ($VH$), and associated production with a top quark pair ($ttH$). Columns represent Higgs boson decay modes. }
	\label{rif}
\end{table}

We can select any part of interest or with high experimental precision to study. For example,
	when we choose the intersection of the first two rows and the first two columns, 
	the determinant of this $2\times2$ matrix is zero if there is only one Higgs resonance,
\begin{equation*}
	O_{g g H, \gamma \gamma} O_{VBF, W W} - O_{VBF, \gamma \gamma} O_{g g H, W W} = 0
\end{equation*}
In general case, the equation should be slightly modified as
\begin{equation}
	(1+\delta) O_{g g H, \gamma \gamma} O_{VBF, W W}-O_{VBF, \gamma \gamma} O_{g g H, W W}=0
\end{equation}
here $\delta$ is a factor, representing the degree of deviation. For clarity, this formula can be transformed as follows,
\begin{equation}
	\frac{O_{VBF, \gamma \gamma} / O_{g g H, \gamma \gamma}}{O_{VBF, W W} / O_{g g H, W W}}= 1 + \delta
\end{equation}
This means that if there is only one Higgs resonance, this type of double ratio is strictly equal to 1, and $\delta = 0$. Whereas, if there are two or more Higgs resonance, the ratio deviates from one, and $\delta \neq 0$. 
Note that the observable $O_{if}$ is defined in Eq.(\ref{R}), and we should sum over the index of the resonance if there are two or more Higgs resonance.
In our work, we focus on the scenario with two mass-degenerate Higgs bosons in the $\mu$NMSSM considering various constraints.


\subsection{Dark Matter in the $\mu$NMSSM}
We have two requirements for DM in the $\mu$NMSSM in our work. Firstly, we suppose that there was a large amount of DM in the early universe, and they reached the current Planck observation $\Omega_{D M} h^{2}=0.120 \pm 0.01$ as they freezed out \cite{37,38,39}.
In this case, the relic density of DM $\tilde{\chi}_{1}^{0}$ in the $\mu$NMSSM is required to be less than the central value 0.12. 
Note that the $\mathbb{Z}_3$-NMSSM employs four parameters: $m_{\tilde{\chi}_{1}^{0}}$, $\lambda$, $\kappa$ and tan$\beta$ to describe the properties of DM,
but the properties of DM in the $\mu$NMSSM are described by five parameters: $m_{\tilde{\chi}_{1}^{0}}$, $\lambda$, $\kappa$, tan$\beta$ and $\mu_{tot}$ \cite{42}.
We can change $\kappa$ to use the $\tilde{\chi}_{1}^{0} \tilde{\chi}_{1}^{0} \rightarrow h_{s} A_{s}$ process to achieve the correct relic density.
Besides, DM based on higgsino in the $\mu$NMSSM has obvious advantages since that higgsino-dominated DM is different from the SU(2) singlet dominated DM which does not interact with Z bosons, such as $
\widetilde{\mathrm{B}}$, $\widetilde{\mathrm{W}}$, and $\widetilde{\mathrm{S}}$.
And the coupling of Z boson with higgsino-dominated DM $C_{Z \tilde{\chi}_{1}^{0} \tilde{\chi}_{1}^{0}}$ in the $\mu$NMSSM is \cite{40,41},
\begin{equation}
	\begin{aligned}
		C_{Z \tilde{\chi}_{1}^{0} \tilde{\chi}_{1}^{0}}=&-\frac{i}{2}\left(g_{1} \sin \theta_{W}+g_{2} \cos \theta_{W}\right)\left(N_{13}^{2}-N_{14}^{2}\right) \gamma_{\mu} P_{L} \\
		&+\frac{i}{2}\left(g_{1} \sin \theta_{W}+g_{2} \cos \theta_{W}\right)\left(N_{13}^{2}-N_{14}^{2}\right) \gamma_{\mu} P_{R}
	\end{aligned}
\end{equation}
where $\left|N_{13}^{2}-N_{14}^{2}\right|$ is called ‘higgsino asymmetry’. 

Secondly, spin-independent (SI) and spin-dependent (SD) DM-nucleon cross sections are required to meet experimental limits \cite{12,12-2,13}.
When the squarks are heavy, the SI nucleon scattering is primarily the t-channel process of exchanging Higgs bosons with the cross-section given as \cite{45,46},
\begin{equation}
	\sigma_{N}^{\mathrm{SI}}=\frac{4 \mu_{r}^{2}}{\pi}\left|f^{(N)}\right|^{2}, \quad f^{(N)}=\sum_{i}^{3} f_{H_{i}}^{(N)}=\sum_{i}^{3} \frac{C_{\tilde{\chi}_{1}^{0} \tilde{\chi}_{1}^{0} H_{i}} C_{N N H_{i}}}{2 m_{H_{i}}^{2}}
	\label{eqUR}
\end{equation}
where $\mu_{r}=m_{N} m_{\tilde{\chi}_{1}^{0}} /\left(m_{N}+m_{\tilde{\chi}_{1}^{0}}\right)$ is the reduced mass of  nucleus and $\tilde{\chi}_{1}^{0}$, and $C_{N N H_{i}}$ is the coupling of Higgs boson  $H_i$ with nucleon,
\begin{equation}
	C_{N N H_{i}}=-\frac{m_{N}}{v}\left[F_{d}^{(N)}\left(V_{i 2}-\tan \beta V_{i 1}\right)+F_{u}^{(N)}\left(V_{i 2}+\frac{1}{\tan \beta} V_{i 1}\right)\right]
\end{equation}
In the above formula, $F_{d}^{(N)}=f_{d}^{(N)}+f_{s}^{(N)}+\frac{2}{27} f_{G}^{(N)}$ and $F_{u}^{(N)}=f_{u}^{(N)}+\frac{4}{27} f_{G}^{(N)}$ with form factor $f_{q}^{(N)}=m_{N}^{-1}\left\langle N\left|m_{q} q \bar{q}\right| N\right\rangle$ ($q=u, d, s$) and $f_{G}^{(N)}=1-\sum_{q=u, d, s} f_{q}^{(N)} .$ 
In our case that mass of the heaviest neutral CP-even Higgs is larger than 1.5~TeV, the dominant contributions to the SI cross-section are from the two light Higgs bosons and the relevant couplings $C_{\tilde{\chi}_{1}^{0} \tilde{\chi}_{1}^{0} H_i}$ is given by \cite{16},
\begin{equation}
	\begin{aligned}
		\label{CXXH}
		C_{\tilde{\chi}_{1}^{0} \tilde{\chi}_{1}^{0}H_i} & \simeq \frac{\mu+\mu_{\mathrm{eff}}}{v}\left(\frac{\lambda v}{\mu+\mu_{\mathrm{eff}}}\right)^{2} \frac{N_{15}^{2} V_{i2}\left(m_{\tilde{\chi}_{1}^{0}} /\left(\mu+\mu_{\mathrm{eff}}\right)-\sin 2 \beta\right)}{1-\left(m_{\tilde{\chi}_{1}^{0}} /\left(\mu+\mu_{\mathrm{eff}}\right)\right)^{2}} \\
		&+\frac{\lambda}{2 \sqrt{2}}\left(\frac{\lambda v}{\mu+\mu_{\mathrm{eff}}}\right)^{2} \frac{N_{15}^{2} V_{i3} \sin 2 \beta}{1-\left(m_{\tilde{\chi}_{1}^{0}} /\left(\mu+\mu_{\mathrm{eff}}\right)\right)^{2}} \\
		&-\sqrt{2} \kappa N_{15}^{2} V_{i3}\left[1+\left(\frac{\lambda v}{\sqrt{2}\left(\mu+\mu_{\mathrm{eff}}\right)}\right)^{2} \frac{1}{1-\left(m_{\tilde{\chi}_{1}^{0}} /\left(\mu+\mu_{\mathrm{eff}}\right)\right)^{2}} \frac{\mu_{\mathrm{eff}}}{\mu+\mu_{\mathrm{eff}}}\right]
	\end{aligned}
\end{equation}
Then the SD cross section takes the following simple from \cite{47,48},
\begin{equation}
	\begin{aligned}
		\sigma_{N}^{\mathrm{SD}} & \simeq C_{N} \times 10^{-4} \times\left(\frac{N_{13}^{2}-N_{14}^{2}}{0.1}\right)^{2} \\
		& \simeq C_{N} \times 10^{-2} \times\left(\frac{\lambda v}{\sqrt{2}\left(\mu+\mu_{\mathrm{eff}}\right)}\right)^{4}\left(\frac{N_{15}^{2} \cos 2 \beta}{1-\left(m_{\tilde{\chi}_{1}^{0}} /\left(\mu+\mu_{\mathrm{eff}}\right)\right)^{2}}\right)^{2}
	\end{aligned}
\end{equation}
where $C_N = C_p  \simeq 4.0$ pb for the proton and $C_N = C_n  \simeq 3.1$ pb for the neutron and these equations could be helpful for understanding our numerical results in the following sections.

\subsection{$(g - 2)_\mu$ in the $\mu$NMSSM}
In supersymmetry, the correction for  the muon anomalous magnetic moment $a_{\mu}^{\mathrm{SUSY}}$ in the $\mu$NMSSM is almost the same as the case in MSSM, the difference is that $\mu$ in MSSM is replaced by $\mu_{tot}$ in the $\mu$NMSSM. 
Although at one loop there is also the possibility of a singlino insertion contribution, 
this contribution is never large when DM constraints are considered, at least for the singlino-dominated DM case \cite{17}.

\begin{figure}[!htb]
	\begin{center}
		\includegraphics[width=0.4\textwidth]{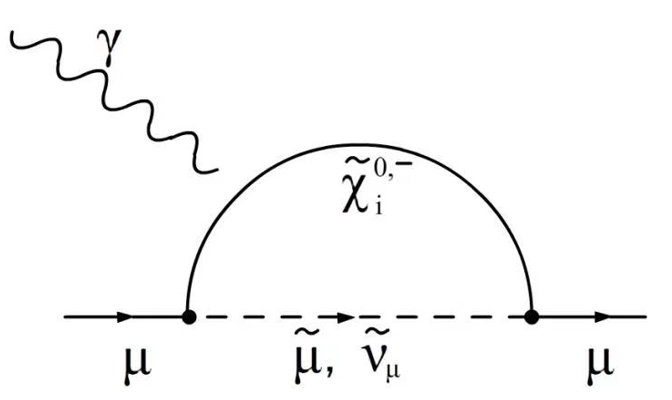}
	\end{center}
	\caption{
		The one-loop contributions arise from Feynman diagrams where the muon lepton number is carried by $\tilde{\mu}$ or
		$\tilde{\nu}_{\mu}$.The external photon can couple to each charged particle.}
	\label{FM1}	
\end{figure}

\begin{figure}[!htb]
	\begin{center}
		\includegraphics[width=1\textwidth]{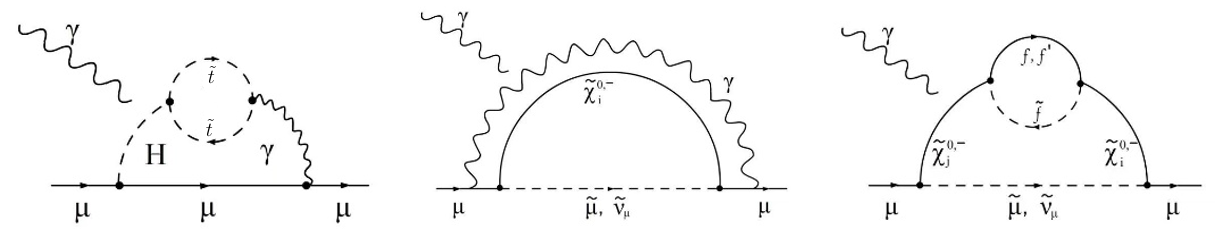}
	\end{center}
	\caption{
		The two-loop $a_{\mu}^{2 \mathrm{L}(\mathrm{a})}$ (left) with closed stop loop inserted into an SM-like one-loop diagram with
		Higgs and photon exchange, $a_{\mu}^{2 \mathrm{L}(\mathrm{b})}$(middle) or $a_{\mu}^{2 \mathrm{L}(\mathrm{c})}$(right) corresponding to SUSY one-loop diagrams with additional photon loop, or with fermion/sfermion-loop
		insertion. The external photon can couple to each charged particle. }
	\label{FM2}
\end{figure}

We list the loop contribution items of $a_{\mu}^{\mathrm{SUSY}}$ \cite{48a, 3a1},
\begin{equation}
	a_{\mu}^{\mathrm{SUSY}}=\left[a_{\mu}^{1 \mathrm{L}}+a_{\mu}^{2 \mathrm{L}(\mathrm{a})}+a_{\mu}^{2 \mathrm{L}(\mathrm{b})}+a_{\mu}^{2 \mathrm{L}(\mathrm{c})}\right]_{\tan{\beta} }+\cdots.
\end{equation}
Referring to Ref.~\cite{li1}, Feynman diagrams for $a_{\mu}^{1 \mathrm{L}}$, $a_{\mu}^{2 \mathrm{L}(\mathrm{a})}$, $a_{\mu}^{2 \mathrm{L}(\mathrm{b})}$, $a_{\mu}^{2 \mathrm{L}(\mathrm{c})}$ are given in Fig.~\ref{FM1} and Fig.~\ref{FM2}, respectively. The subscript $\tan{\beta}$ means that each n-loop term is proportional to the $(\tan{\beta})^n$ term, which leads to a 10$\%$ correction for higher-order terms if $\tan{\beta}$ is greater than 50 \cite{li11}. If $\tan{\beta}$ tends to infinity, even if all SUSY particle masses are above TeV scale, there can keep a large correction \cite{li35}.

It can be seen from Fig.~\ref{FM1} that the correction of one-loop diagram mainly comes from two parts, one part is the exchange of neutralinos ${\tilde{\chi}^{0}}_i$ and smuon $\tilde{\mu}$, and the other part is the exchange of charginos ${\tilde{\chi}^{\pm}}_i$ and sneutrino $\tilde{\nu}_{\mu}$:

\begin{equation}
	\begin{gathered}
		a_{\mu}^{\mathrm{1L}}=a_{\mu}^{\tilde{\chi}^{0} \tilde{\mu}}+a_{\mu}^{\tilde{\chi}^{\pm} \tilde{\nu}} \\
		a_{\mu}^{\tilde{\chi}^{0} \tilde{\mu}}=\frac{m_{\mu}}{16 \pi^{2}} \sum_{i, l}\left\{-\frac{m_{\mu}}{12 m_{\tilde{\mu}_{l}}^{2}}\left(\left|n_{i l}^{\mathrm{L}}\right|^{2}+\left|n_{i l}^{\mathrm{R}}\right|^{2}\right) F_{1}^{\mathrm{N}}\left(x_{i l}\right)+\frac{m_{\tilde{\chi}_{i}^{0}}}{3 m_{\tilde{\mu}_{l}}^{2}} \operatorname{Re}\left(n_{i l}^{\mathrm{L}} n_{i l}^{\mathrm{R}}\right) F_{2}^{\mathrm{N}}\left(x_{i l}\right)\right\}, \\
		a_{\mu}^{\tilde{\chi}^{\pm} \tilde{\nu}}=\frac{m_{\mu}}{16 \pi^{2}} \sum_{k}\left\{\frac{m_{\mu}}{12 m_{\tilde{\nu}_{\mu}}^{2}}\left(\left|c_{k}^{\mathrm{L}}\right|^{2}+\left|c_{k}^{\mathrm{R}}\right|^{2}\right) F_{1}^{\mathrm{C}}\left(x_{k}\right)+\frac{2 m_{\tilde{\chi}_{k}^{\pm}}}{3 m_{\tilde{\nu}_{\mu}}^{2}} \operatorname{Re}\left(c_{k}^{\mathrm{L}} c_{k}^{\mathrm{R}}\right) F_{2}^{\mathrm{C}}\left(x_{k}\right)\right\},
	\end{gathered}
\end{equation}
where $i$ = 1,...,5, $l$ = 1, 2, $k$ = 1, 2 are the neutralino, smuon and chargino index, respectively. $n_{i l}^{\mathrm{L}}$, $n_{i l}^{\mathrm{R}}$, $c_{k}^{\mathrm{L}}$, $c_{k}^{\mathrm{R}}$ are expressed as follows,

\begin{equation}
	\begin{gathered}
		n_{i l}^{\mathrm{L}}=\frac{1}{\sqrt{2}}\left(g_{2} N_{i 2}+g_{1} N_{i 1}\right) X_{l 1}^{*}-y_{\mu} N_{i 3} X_{l 2}^{*}, \quad n_{i l}^{\mathrm{R}}=\sqrt{2} g_{1} N_{i 1} X_{l 2}+y_{\mu} N_{i 3} X_{l 1}, \\
		c_{k}^{\mathrm{L}}=-g_{2} V_{k 1}^{\mathrm{c}}, \quad c_{k}^{\mathrm{R}}=y_{\mu} U_{k 2}^{\mathrm{c}} .
	\end{gathered}
\end{equation}
where $y_{\mu}=g_{2} m_{\mu} /( \sqrt{2} m_{W} \cos \beta)$ is the muon Yukawa coupling. $X$ is the  smuon mass rotation matrix. The kinematic loop functions $F(x)$ depending on the variables  $x_{i l} \equiv m_{\tilde{\chi}_{i}^{0}}^{2} / m_{\tilde{\mu}_{l}}^{2}$ and $x_{k} \equiv m_{\tilde{\chi}_{k}^{\pm}}^{2} / m_{\tilde{\nu}_{\mu}}^{2}$ are given by:

\begin{equation}
	\begin{aligned}
		&F_{1}^{N}(x)=\frac{2}{(1-x)^{4}}\left[1-6 x+3 x^{2}+2 x^{3}-6 x^{2} \ln x\right] \\
		&F_{2}^{N}(x)=\frac{3}{(1-x)^{3}}\left[1-x^{2}+2 x \ln x\right] \\
		&F_{1}^{C}(x)=\frac{2}{(1-x)^{4}}\left[2+3 x-6 x^{2}+x^{3}+6 x \ln x\right] \\
		&F_{2}^{C}(x)=-\frac{3}{2(1-x)^{3}}\left[3-4 x+x^{2}+2 \ln x\right]
	\end{aligned}
\end{equation}

It can be seen that $a_{\mu}^{\mathrm{1L}}$ depends essentially on the bino(wino) masses $M_1$($M_2$), the higgsino mass $\mu$, the left(right) smuon mass  $m_{L_{\mu}}$, $m_{E_{\mu}}$ and $\tan\beta$, and it hardly depends on $A_t$,$A_\kappa$ or $A_\lambda$.
There is a simple relationship, $a_{\mu}^{\mathrm{1L}}$ is proportional to $\tan{\beta}/M_{SUSY}^{2}$, where $M_{SUSY}^{2}$ refers to the general SUSY mass \cite{li13,li15,li88}.

The two-loop correction to $a_{\mu}^{\mathrm{SUSY}}$ has three types shown in Fig.~\ref{FM2}. 
The correction from the left diagram mainly depends on Higgs boson masses and stop masses. 
The term $a_{\mu}^{2 \mathrm{L}(\mathrm{a})}$ can be large in certain regions of parameter space but decouples as the masses of SUSY particles or heavy Higgs increase \cite{li1}.
The term $a_{\mu}^{2 \mathrm{L}(\mathrm{b})}$ has an additional negative factor, which usually results in a (-7...-9)$\%$ correction according to the literature\cite{li12,48k}. The contribution of the fermion/sfermion-loop is introduced in the right diagram, which includes the  each generation squarks and sleptons. The peculiarity of this diagram is that if the mass of the fermions (sfermions) is very large, it does not decouple, but increases logarithmically. When the squark mass is lower than the TeV scale, the $a_{\mu}^{2 \mathrm{L}(\mathrm{c})}$ term can achieve a positive or negative correction of about 10$\%$ \cite{li13,li14}.

\subsection{Parameter space and scanning method}
We start with a random scan to find the initial point in the parameter space, and then use the Markov chain Monte Carlo (MCMC) scan in EasyScan-1.0.0 \cite{49} to explore the high-dimensional parameter space as follows,
\begin{equation}
	\begin{array}{c}
	\left|M_{1}\right| \leq 2.5\ \text{TeV}, \quad 100\ \text{GeV} \leq M_{2} \leq 2\ \text{TeV}, \\
		0 \leq \lambda \leq 0.5, \quad|\kappa| \leq 0.5, \quad 1 \leq \tan \beta \leq 60, \quad 2\ \text{TeV} \leq \left|A_{t}\right| \leq 5\ \text{TeV}, \\
		10\ \text{GeV} \leq \mu \leq 1\ \text{TeV}, \quad 100\ \text{GeV} \leq \mu_{\mathrm{tot}} \leq 1\ \text{TeV}, \quad\left|A_{\kappa}\right| \leq 1\ \text{TeV}, \\
		100\ \text{GeV} \leq m_{L_{\mu}} \leq 2\ \text{TeV}, \quad 100\ \text{GeV} \leq m_{E_{\mu}} \leq 3\ \text{TeV}, 
	\end{array}
\end{equation}
We set $A_\lambda$ and other supersymmetric parameters of the first and third generation sleptons, all squarks and gluinos to 2~TeV.
We use SARAH-4.14.3 \cite{50,51,52,53} to generate the model files in the $\mu$NMSSM and then use SPheno-4.0.4 \cite{54,55} to generate spectrums, which includes the masses of the particles, mixing angles between mass eigenstates and interaction fields and other physical observables. 
In order to improve the sampling efficiency, we only require that the spectrums should satisfy the following conditions during scanning, whereas we will discuss the effects of DM direct detection limits and the $(g-2)_\mu$ anomaly on the parameter space in Section 3 and Section 4, respectively.
\begin{itemize}
	\item
There are two mass-generate Higgs bosons with mass  in 123-128~GeV. 
The package HiggsBounds-5.3.2\cite{57} is used to check non-SM-like Higgs bosons against the published exclusion bounds from Higgs boson searches at the LEP, Tevatron and LHC experiments.
While the package HiggsSignals-2.2.3\cite{56} evaluates a $\chi^2$ measure to provide a quantitative answer to the statistical question of how compatible the Higgs data is with predictions.
Spectrums with p-values less than 0.05 are excluded at the 95\% confidence level.
SPheno writes the spectrum file and the input interface files for HiggsBounds and HiggsSignals
	\footnote{The interface files include BR\_t.dat, BR\_H\_NP.dat, BR\_Hplus.dat, effC.dat, MH\_GammaTot.dat, MHplus\_GammaTot.dat.
With these files as input, we can run HiggsBounds and HiggsSignals by issuing the following commands, respectively:
$$ \text{./HiggsBounds\ LandH\ effC\ 5\ 1}  $$
“LandH"  indicates that LEP, Tevatron and LHC analyses will be considered by HiggsBounds, “effC" stands for the input format for HiggsBounds, “5" stands for the number of neutral Higgs bosons and “1" stands for the number of charged Higgs boson in the model.
$$ \text{./HiggsSignals\ latestresults\ peak\ 2\ effC\ 5\ 1}  $$	
“peak" specifies that the peak-centered $\chi^2$ method should be used by HiggsSignals, “2" selects the Gaussian parametrization
for the Higgs mass uncertainty. “latestresults" has the same meaning as “LandH", “effC", “5" and “1" have the same meaning as described for HiggsBounds.}.

 \item 
	DM relic density $\Omega h^{2}=0.120\pm 0.01$. We assume the LSP $\chi_1^0$ is one of the DM candidates, so the DM relic density is required to be less than 0.120, i.e., $\Omega h^{2}\leq0.120$. The Spin-independent (SI) DM cross section $\sigma_{SI}$ and spin-dependent (SD) cross section $\sigma_{SD}$ should be scaled by a factor $\Omega h^{2}/0.120$. We use the package MicrOMEGAs-5.0.4\cite{58,59,60,61,62,63} to calculate $\Omega h^{2}$, $\sigma_{SI}$ and $\sigma_{SD}$.
	\item
		Results from sparticle searches at the LHC. SModelS-1.2.3\cite{68,69,70,71,72,73} is used to determine whether a sample is excluded or not by decomposing spectrum and converting it into Simplified Model topologies and then comparing it with these simplified model results interpreting from the LHC. 
		We consider these typical processes $p p \to \tilde{\chi}_{1,2}^0 \tilde{\chi}_1^\pm,\tilde{\chi}_1^+ \tilde{\chi}_1^-, \tilde{\mu}^+ \tilde{\mu}^- $ and use the package Prospino2\cite{64,65,66,67} to generate the next-to-leading order cross sections of these processes as inputs for SModels.
\end{itemize}

\section{\label{sec:DM}Properties of DM confronted with direct detection limits}

\begin{figure}[!htb]
	\begin{center}
		\includegraphics[width=1\textwidth]{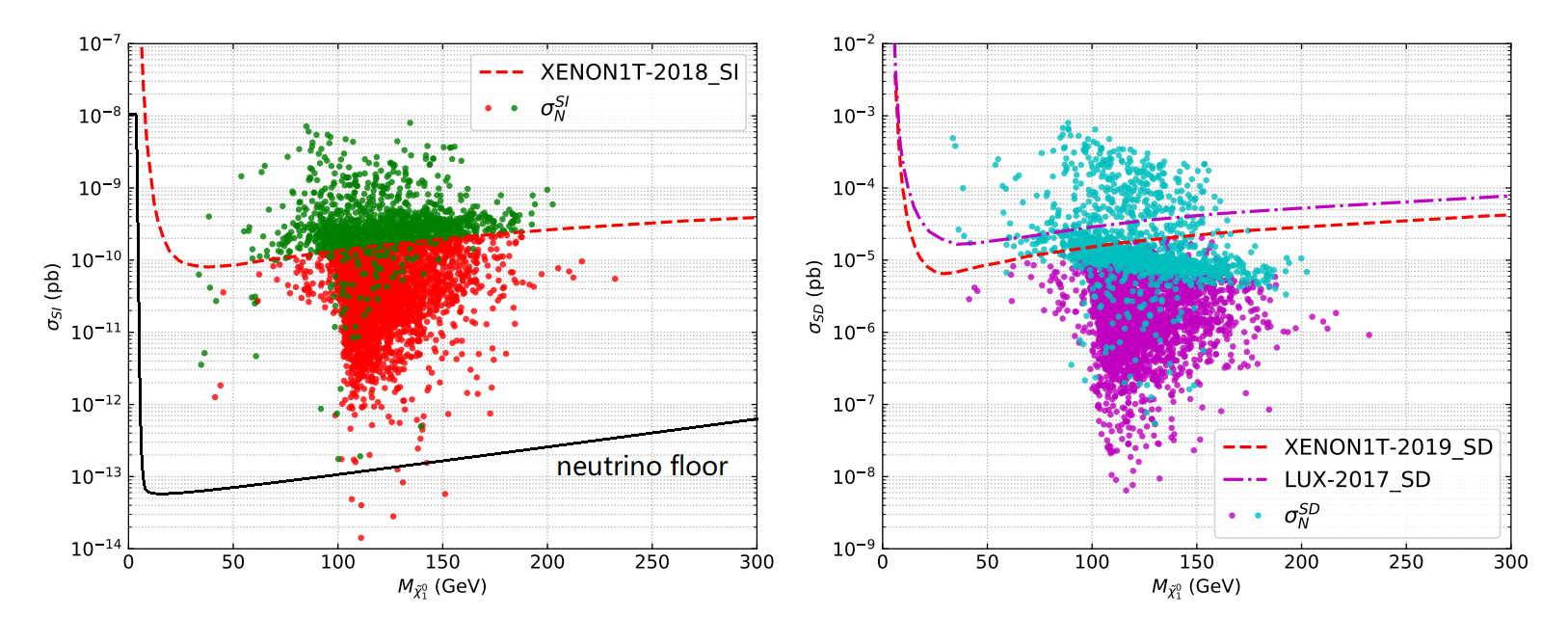}
	\end{center}
	\caption{
SI (left plot) and SD (right plot) cross section for DM-nucleon experiments versus the mass of DM  in the scenario with two mass-degenerate Higgs bosons. The red dashed line on the left plot stands for the XENON-1T SI limit in 2018, the red dashed line (magenta dash-dotted) on the right plot stands for the XENON-1T SD limit in 2019 (LUX SD limit in 2017). Samples above these lines are excluded, and samples with red and magenta points are the surviving samples considering the limits from DM direct detections. The black solid line stands for the neutrino floor.
}
	\label{DM0}
\end{figure}

We project the surviving samples obtained during scanning in Fig. \ref{DM0}.
In the left plot, 
the green samples above the red dashed line are excluded by the XENON-1T SI limit in 2018 \cite{12} and the green samples below the red dashed line are excluded by the XENON-1T SD limit in 2019 \cite{12-2}. 
While in the right plot, the cyan samples above the red dashed line are excluded by the XENON-1T SD limit in 2019, but the cyan samples below the red dashed line are excluded by the XENON-1T SI limit in 2018.
The samples with red points in the left plot and magenta points in the right plot are the surviving samples considering the limits from DM direct detections.
So, we can see that the SI and SD constraints are complementary in limiting the parameter space of $\mu$NMSSM.  
We notice that there are a few of samples whose SI cross section
	can be lower than the neutrino floor, and consequently these
	DM may never be probed in DM direct detections.
In the following discussion, we will focus on the parameter space tightly limited by DM direct detection limits in the $\mu$NMSSM.

\begin{figure}[!htb]
	\begin{center}
		 \includegraphics[width=1\textwidth]{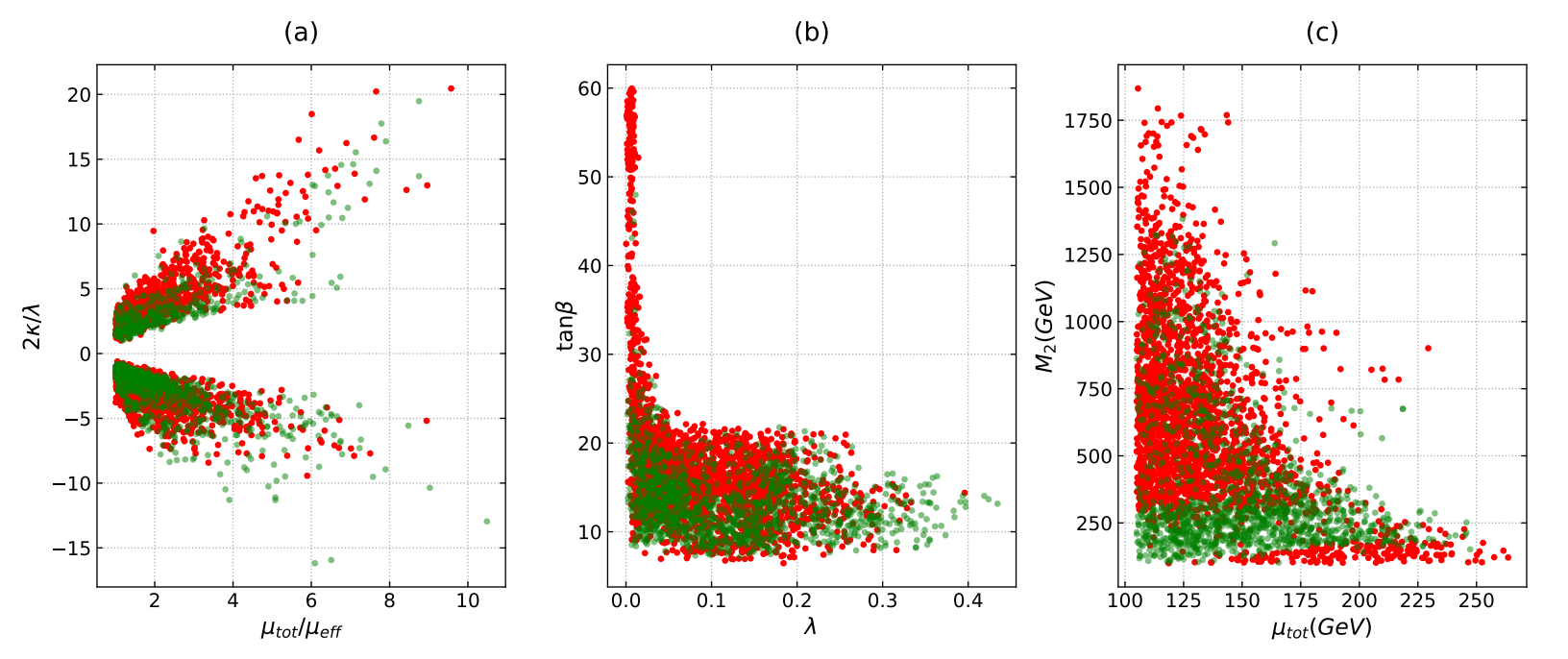}
	\end{center}
	\caption{
	The projection of samples on the $\lambda$ – $\kappa$ plane, $\lambda$ - tan$\beta$ plane and $\mu_{tot}$ - $M_2$ plane. Samples represented by green points indicate ones excluded by DM direct detection limits and those represented by red points are surviving samples confronted with these limits.}
	\label{DM1}
\end{figure}

Fig.~\ref{DM1} shows the samples on the $2\kappa/\lambda$ - $\mu_{tot}/\mu_{eff}$ plane, $\lambda$ - tan$\beta$ plane and $\mu_{tot}$ - $M_2$ plane.  Samples with red (green) points are allowed (excluded) by DM direct detection limits. In the scenario with two mass-degenerate SM-like Higgs bosons, the DM is higgsino-dominated, which is easier to meet the constraints from DM direct detections compared with the scenario with singlino-dominated DM.
To avoid the singlino-dominated DM in the $\mu$NMSSM, $2\kappa/\lambda$ should be far greater than $\mu_{tot}/\mu_{eff}$, which can be seen clearly from Fig.~\ref{DM1}(a).
This is significantly different from the case in the $\mathbb{Z}_3$-NMSSM, which only requires $2\kappa/\lambda$ to be greater than 1.
In Fig.~\ref{DM1}(b), we find that a large tan$\beta$ greater than about 20 is accompanied with a small $\lambda$ less than about 0.05, which guarantee to realize two mass-degenerate Higgs boson with mass about 125 GeV. In Fig.~\ref{DM1}(c), we can see that a larger $M_2$ is required since that it can avoid the wino-dominated neutralino as the LSP.

\begin{figure}[!htb]
	\begin{center}
		\includegraphics[width=0.9\textwidth]{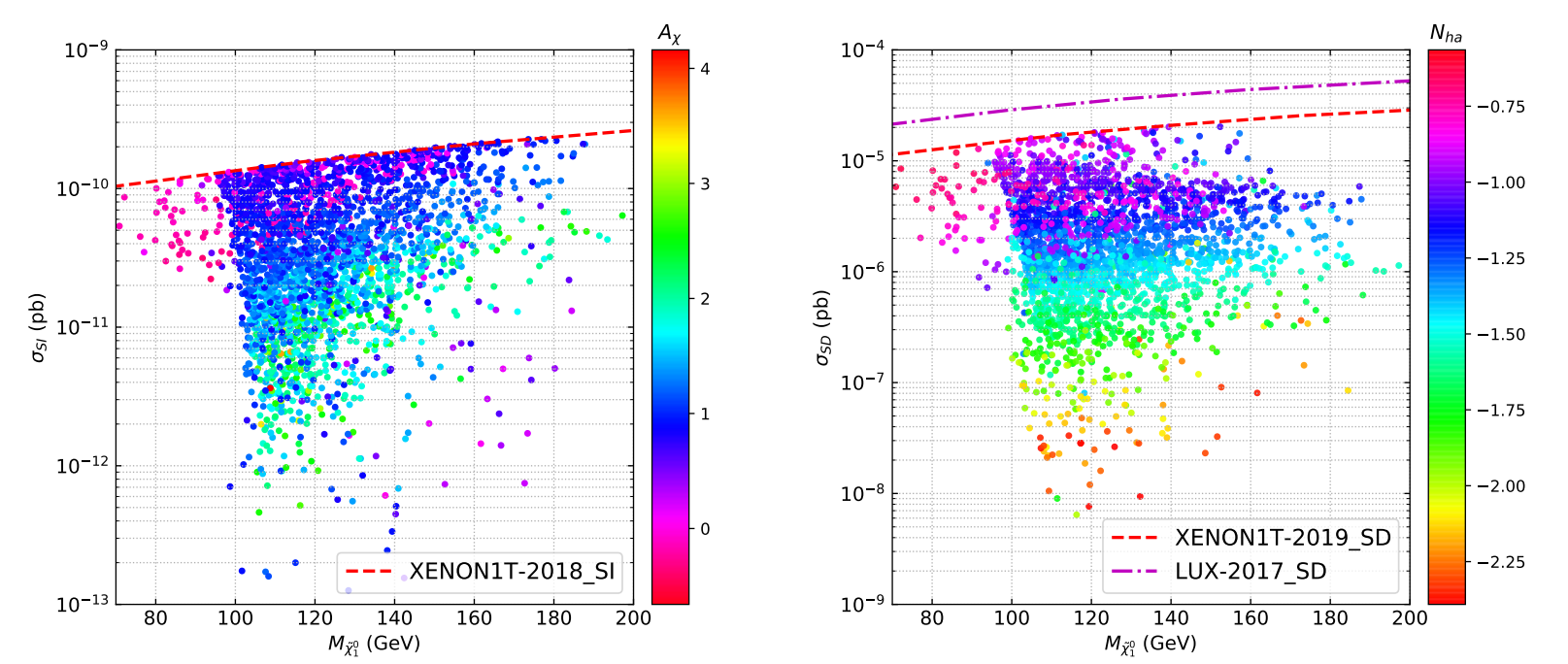}
	\end{center}
	\caption{	
		The samples within the DM direct detection limits in Fig.~\ref{DM0}. The colorbar represents $A_X$  ($N_{ha}$) on the left (right) plot. We use the logarithm of $A_X$ and $N_{ha}$ to reduce the impact of extreme values on the color distribution.}
	\label{DM3}
\end{figure}

To consider the DM direct detection limits, we try to find the factor that affect the variation of SI and SD cross section. We find that  $m_{\tilde{\chi}_{1}^{0}}$ is  around 50-200 GeV,  which is much greater than the neutron mass. So $\mu_r$ in  Eq.(\ref{eqUR})  change little for different $m_{\tilde{\chi}_{1}^{0}}$, $\sigma_{SI}$ is primarily influenced by the couplings $C_{\tilde{\chi}_{1}^{0} \tilde{\chi}_{1}^{0} H_i}$ and $C_{N N H_{i}}$. 
We propose that the factor $A_{X}$ derived from Eq.~(\ref{CXXH}) has a powerful influence on $\sigma_{SI}$:
\begin{equation}
	\label{AX}
	A_{X}=\frac{{m_{\widetilde{\chi}_{1}^{0} }/ \mathbf{\mu}_{t o t}}}{1-\left(m_{\widetilde{\chi}_{1}^{0}} / \mathbf{\mu}_{t o t}\right)^{2}}
\end{equation}
In Fig.~\ref{DM3} we show the effect of $A_X$ on $\sigma_{SI}$ on the left plot, and the effect of higgsino asymmetry $N_{ha}=\left|N_{13}^{2}-N_{14}^{2}\right|$ on $\sigma_{SD}$ on the right plot. As the size of the cross section decreases, the color of the surviving sample shows a gradient change, which shows that the SI and SD cross sections depend on the variation of $A_X$ and $N_{ha}$.

\begin{figure}[!htb]
	\begin{center}
		\includegraphics[width=0.9\textwidth]{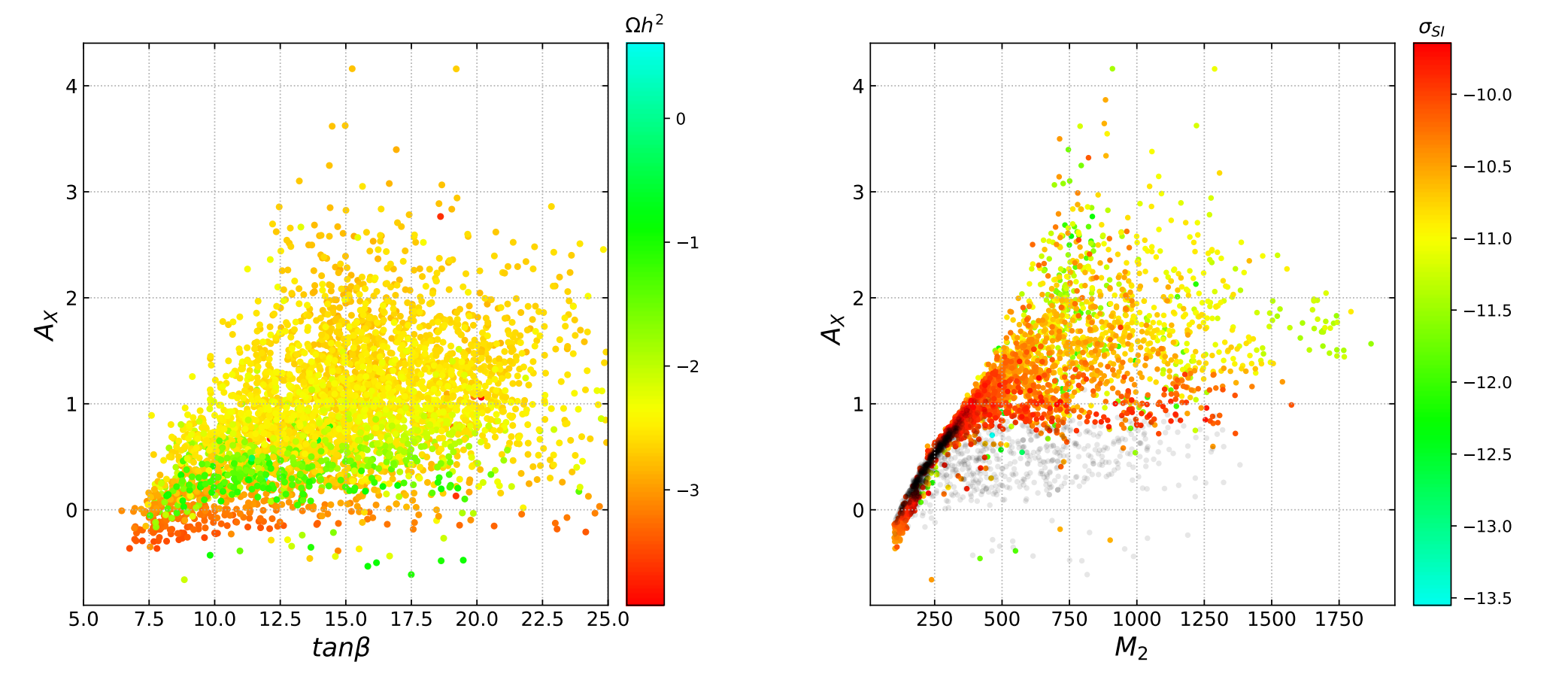}
	\end{center}
	\caption{
		All samples in Fig.~\ref{DM0} are projected on $\tan\beta$ - $A_X$ plot and $M_2$ - $A_X$ plot. The colorbar represents the common logarithm of DM relic density in the left plot. 
		Black samples in the right plot are excluded by DM direct detection limits, and the others are surviving samples. For surviving samples, the colorbar in the right plot represents the common logarithm of $\sigma_{SI}$, samples close to green color are much safer, samples close to red color are near DM direct detection limits.}
	\label{DM5}
\end{figure}

We project samples on $\tan\beta$ - $A_X$ plot and $M_2$ - $A_X$ plot in Fig.\ref{DM5}.
The trend that $\sigma_{SI}$ decreases with increasing $A_X$ holds true for all samples as shown on the right plot. In the horizontal band $A_X=0\sim1$, a large number of samples are excluded because they have higher DM relic densities leading to higher $\sigma_{SD}$ and $\sigma_{SI}$. This can be seen from the green samples in the horizontal band $A_X=0\sim1$ on the left plot. $\Omega h^2$ of samples in this band are much higher than the surrounding ones and so they are excluded by DM direct detection limits.

\begin{figure}[!htb]
	\begin{center}
		\includegraphics[width=0.9\textwidth]{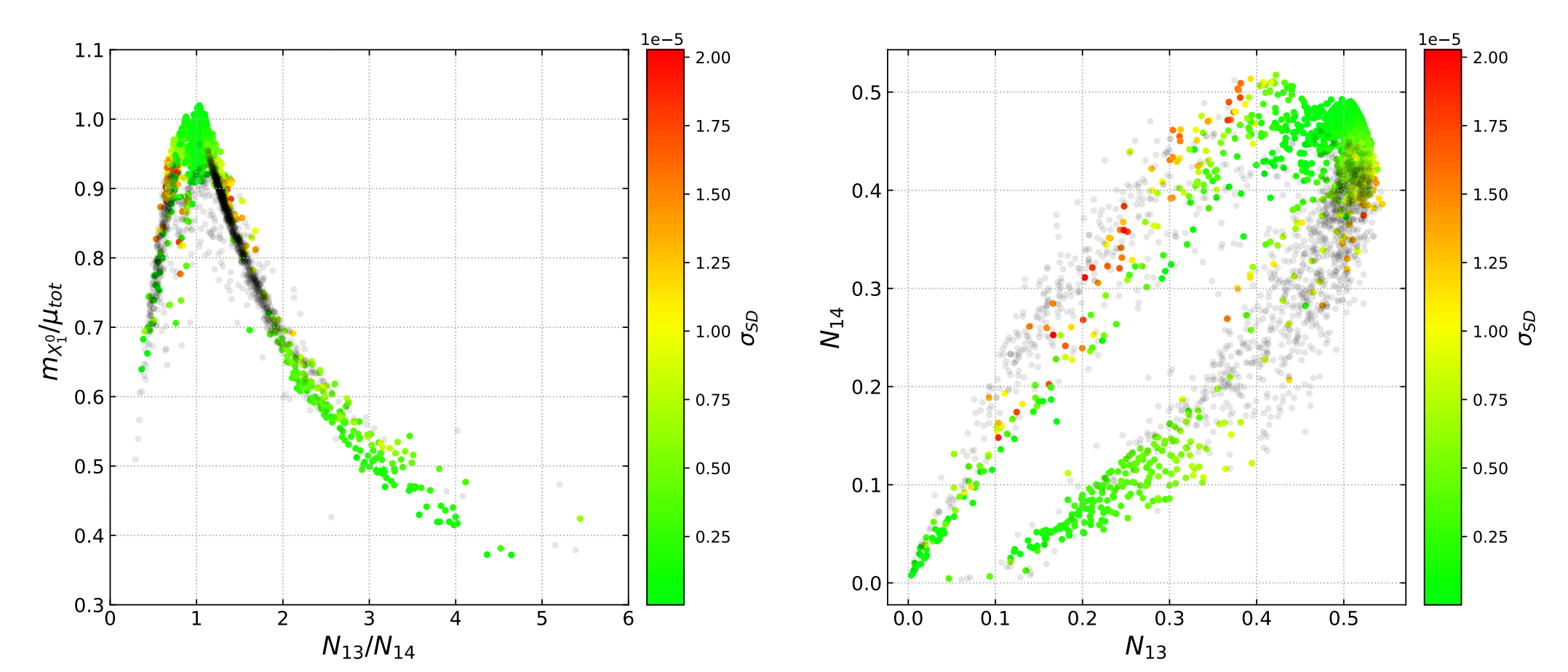}
	\end{center}
	\caption{
	The projection of all samples in Fig.~\ref{DM0} on $N_{13}/N_{14}$ - $m_{\widetilde{\chi}_{1}^{0}} / \mu_{tot}$ plot and $N_{13}$ - $N_{14}$ plot. Black samples in each plot are excluded by DM direct detection limits, and the others are surviving samples. For surviving samples, the colorbar in each plot represents $\sigma_{SD}$, samples close to green color are much safer, samples close to red color are near DM direct detection limits. }
		\label{DM4}
	\end{figure}

We project samples on $N_{13}/N_{14}$ - $m_{\widetilde{\chi}_{1}^{0}} / \mu_{tot}$ plot and $N_{13}$-$N_{14}$ plot in Fig.\ref{DM4}.
When $N_{13}$ is less than $N_{14}$, $m_{\widetilde{\chi}_{1}^{0}} / \mu_{tot}$ is proportional to $N_{13}$/$N_{14}$. 
When $N_{13}$ is relatively large, however, the relationship between $m_{\widetilde{\chi}_{1}^{0}} / \mu_{tot}$ and $N_{13}$/$N_{14}$ is inversely proportional.
We can see that the samples with $N_{13}$/$N_{14}$ close to one on the left plot of Fig.~\ref{DM4} correspond to the samples with smaller higgsino asymmetry on the right plot of Fig.~\ref{DM3}, and they all have safer $\sigma_{SI}$.
The horizontal band $m_{\widetilde{\chi}_{1}^{0}} / \mu_{tot}= 0.7 - 0.9$ is severely excluded by DM direct detection limits because it corresponds to a larger higgsino asymmetry.
When $N_{13}$/$N_{14}$ is smaller or larger, the surviving samples reappear because $N_{13}$ and $N_{14}$ themselves are much small.
This corresponds to the green samples in the left-bottom corner on the right plot in Fig.~\ref{DM4}. 
The right plot supports our conclusion that the samples satisfying DM direct detection limits must have a smaller higgsino asymmetry.

\section{\label{TH}The Mass-degenerate SM-like Higgs bosons and the explanation of  Muon $g-2$  Anomaly}

\subsection{The Mass-degenerate  SM-like Higgs bosons}

To show the deviation of Higgs signal ($h$) observed by the ATLAS and CMS Collaborations from the SM prediction, we define  
\begin{equation}
	R_{if}^{h_{\alpha}} \equiv \frac{\sigma\left(i \rightarrow H_{\alpha}\right) \times \mathrm{B}\left(H_{\alpha} \rightarrow f\right)}{\sigma\left(i \rightarrow h_{\mathrm{SM}}\right) \times \mathrm{B}\left(h_{\mathrm{SM}} \rightarrow f\right)},\\
~~~~	\mu_i^{H_\alpha}\equiv\frac{\sigma_i^{H_\alpha}}{\sigma_i^{h_{SM}}}= \frac{\sigma\left(i \rightarrow H_{\alpha}\right)}{\sigma\left(i \rightarrow h_{\mathrm{SM}}\right)}
\end{equation}
\begin{equation}
B_{bb/WW}=\sum_{\alpha}[B(H_\alpha \to bb/WW)\times \sum_{i}\sigma_i^{H_\alpha}/\sum_{i}\sigma_i^{h}]
\end{equation}
where $\alpha=1,2$, $i$ denotes the production modes: gluon fusion ($ggH$), vector boson fusion ($VBF$), associated production with a Z or W boson ($VH$), and associated production with a top quark pair ($ttH$), $f$ denotes the decay modes: $\gamma\gamma$, $WW$, $\tau\tau$, $bb$, etc. We can simply have $\sigma_{i}^{h}=\sigma_{i}^{H_{1}}+\sigma_{i}^{H_{2}}$ and $R_{i f}^{h}=R_{i f}^{H_{1}}+R_{i f}^{H_{2}}$. In the calculations we ignore the interference effect between the two mass degenerate Higgs bosons in the $\mu$NMSSM because the mass difference between most of the two Higgs bosons in our collected samples is greater than 50 MeV. 

\begin{figure}[!htb]
	\begin{center}
		\includegraphics[width=1\textwidth]{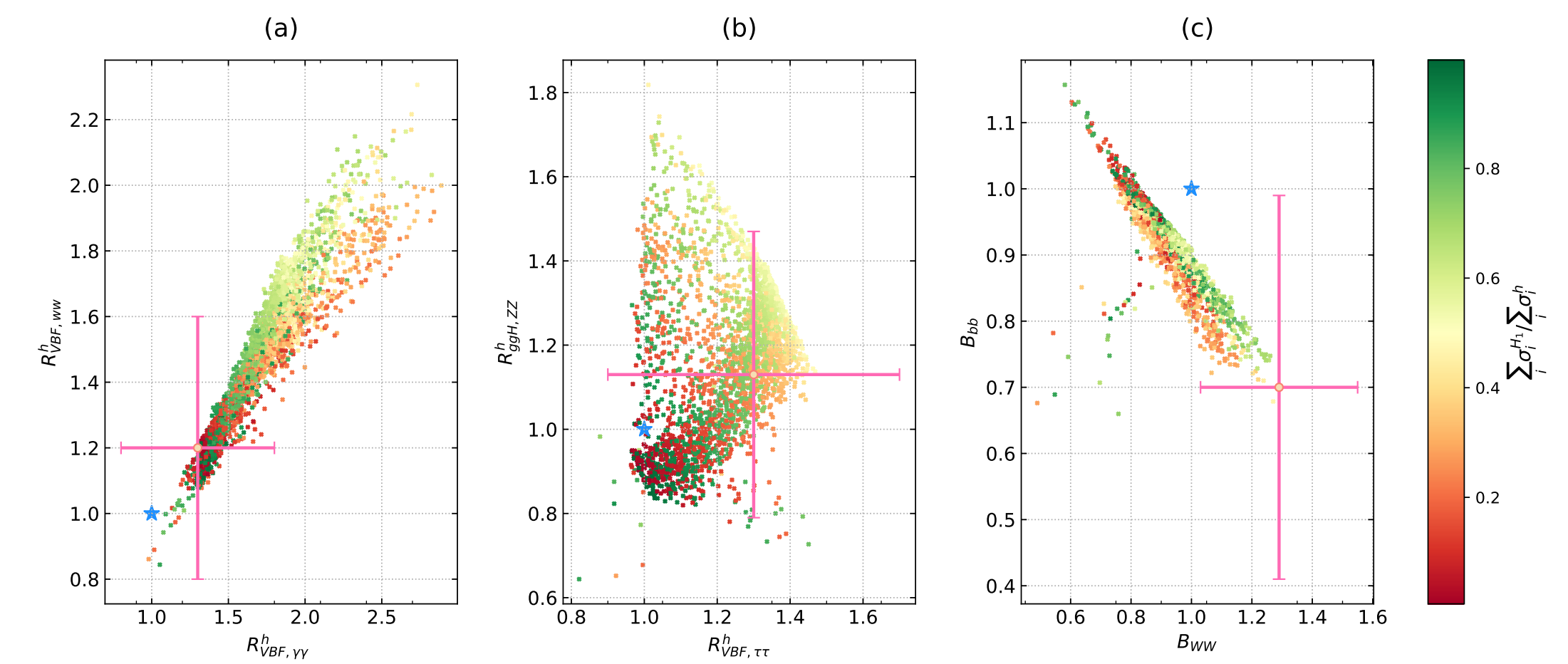}
	\end{center}
	\caption{
		All surviving samples in Fig.~\ref{DM3}  are projected on plots of $R^h_{VBF,\gamma\gamma}$ - $R^h_{VBF,WW}$, $R^h_{VBF,\tau\tau}$ - $R^h_{ggH,ZZ}$, $B_{WW}$ - $B_{bb}$. The colorbar represents $\sum_{i}\sigma_i^{H_1}/\sum_{i}\sigma_i^{h}$. 
		Samples with green (red) color indicate that $\sigma_{i}^h$ mainly comes from the contribution of $H_1$ ($H_2$), and samples with golden color indicate that $\sigma^h$ comes from contributions of both $H_1$ and $H_2$. The pink line stands for one standard deviation observed by ATLAS and CMS. The blue star represents the SM prediction. }
	\label{DH5}
\end{figure}

We project the surviving samples on plots of $R^h_{VBF,\gamma\gamma}$ - $R^h_{VBF,WW}$, $R^h_{VBF,\tau\tau}$ - $R^h_{ggH,ZZ}$, $B_{WW}$ - $B_{bb}$ in Fig.~\ref{DH5}. From Fig.~\ref{DH5}, we can see that $R^h$ and $B_{bb(WW)}$ predicted by SM are different from that measured by ATLAS and CMS.  $B_{bb(WW)}$ predicted by SM even deviates beyond the experimental $1\sigma$ error.
But Samples in the scenario with two mass-degenerate Higgs bosons are much more consistent with measured data than the SM prediction. 

To quantify the degree of agreement with the measured data totally, we define
\begin{equation}
	\chi_{i f}^{2}=\frac{1}{2}\left(\frac{R^h_{i f}-R_{i f}^{h, o b}}{R_{i f}^{h, e r}}\right)^{2}
\end{equation}
where $R_{if}^{h, ob}$ is the measured central value and $R_{if}^{h, er}$ is one standard deviation.
When $\chi^2$ is zero, samples are perfectly aligned with measured data. 
Besides, to compare with the SM prediction conveniently, we define
$C_{i f}=\chi_{i f}^{2} /\left(\chi_{i f}^{2}\right)_{S M}$.
If $C_{if}$ is smaller than one,  the scenario with two mass-degenerate Higgs bosons is more consistent with the measured data.
Referring to Ref.~\cite{34}, we also use the double ratios:
\begin{equation}
	\label{D123}
		D_{1} =\frac{R_{\mathrm{VBF,\tau \tau}}^{h} / R_{g gH, \tau \tau}^{h}}{R_{\mathrm{VBF, bb}}^{h} / R_{g gH, bb}^{h}}, 
		D_{2} =\frac{R_{\mathrm{VBF, \gamma \gamma}}^{h} / R_{g gH, \gamma \gamma}^{h}}{R_{\mathrm{VBF, WW}}^{h} / R_{g gH, WW}^{h}}, 
		D_{3} =\frac{R_{\mathrm{VBF,WW}}^{h} / R_{g gH, WW}^{h}}{R_{\mathrm{VBF, bb}}^{h} / R_{g gH, bb}^{h}}
\end{equation}

\begin{figure}[!htb]
	\begin{center}
		\includegraphics[width=1\textwidth]{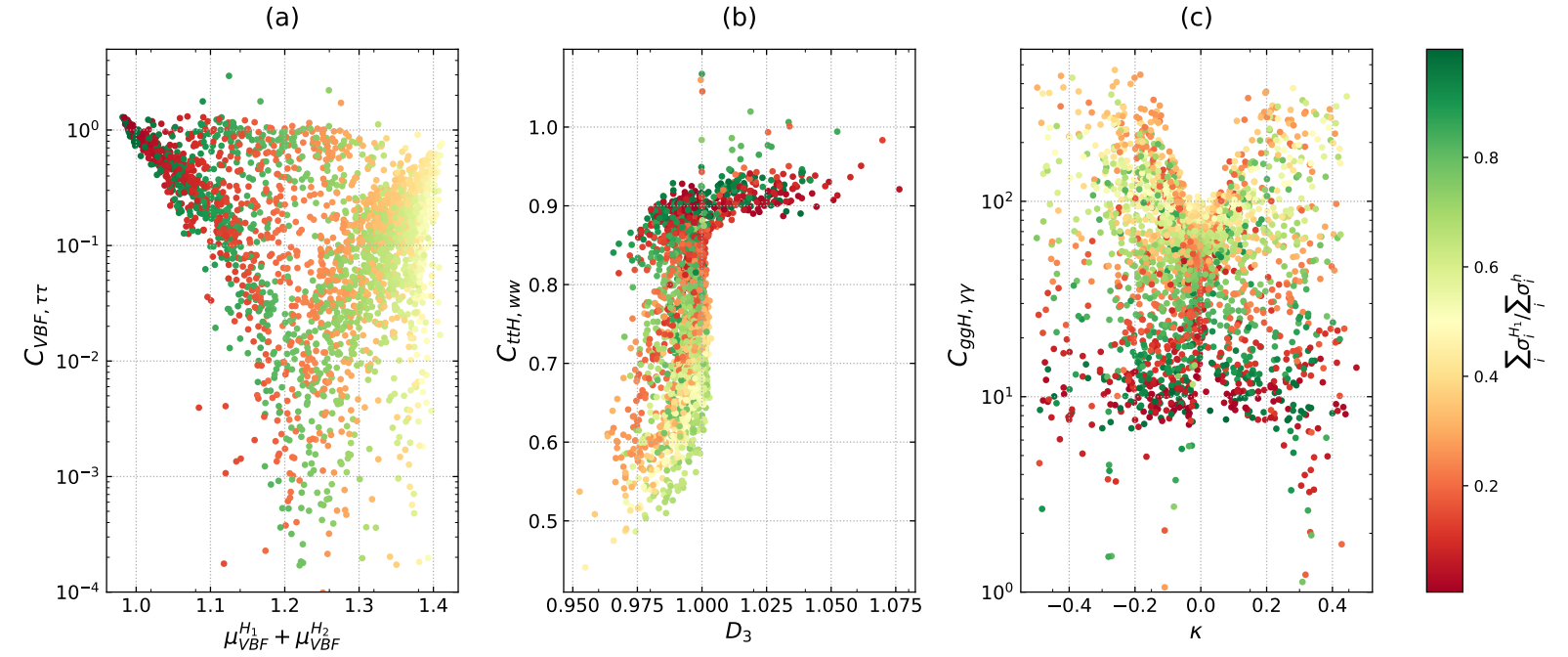}
	\end{center}
	\caption{
		All surviving samples in Fig.~\ref{DM3}  are projected on plots $\mu^{H_1}_{VBF}+\mu^{H_2}_{VBF}$ - $C_{VBF,\tau\tau}$, $D_3$ - $C_{ttH,WW}$ and $\kappa$ - $C_{ggH,\gamma\gamma}$. The colorbar is the same as in Fig.~\ref{DH5}}
	\label{DH6}
\end{figure}

\begin{figure}[!htb]
	\begin{center}
		\includegraphics[width=0.8\textwidth]{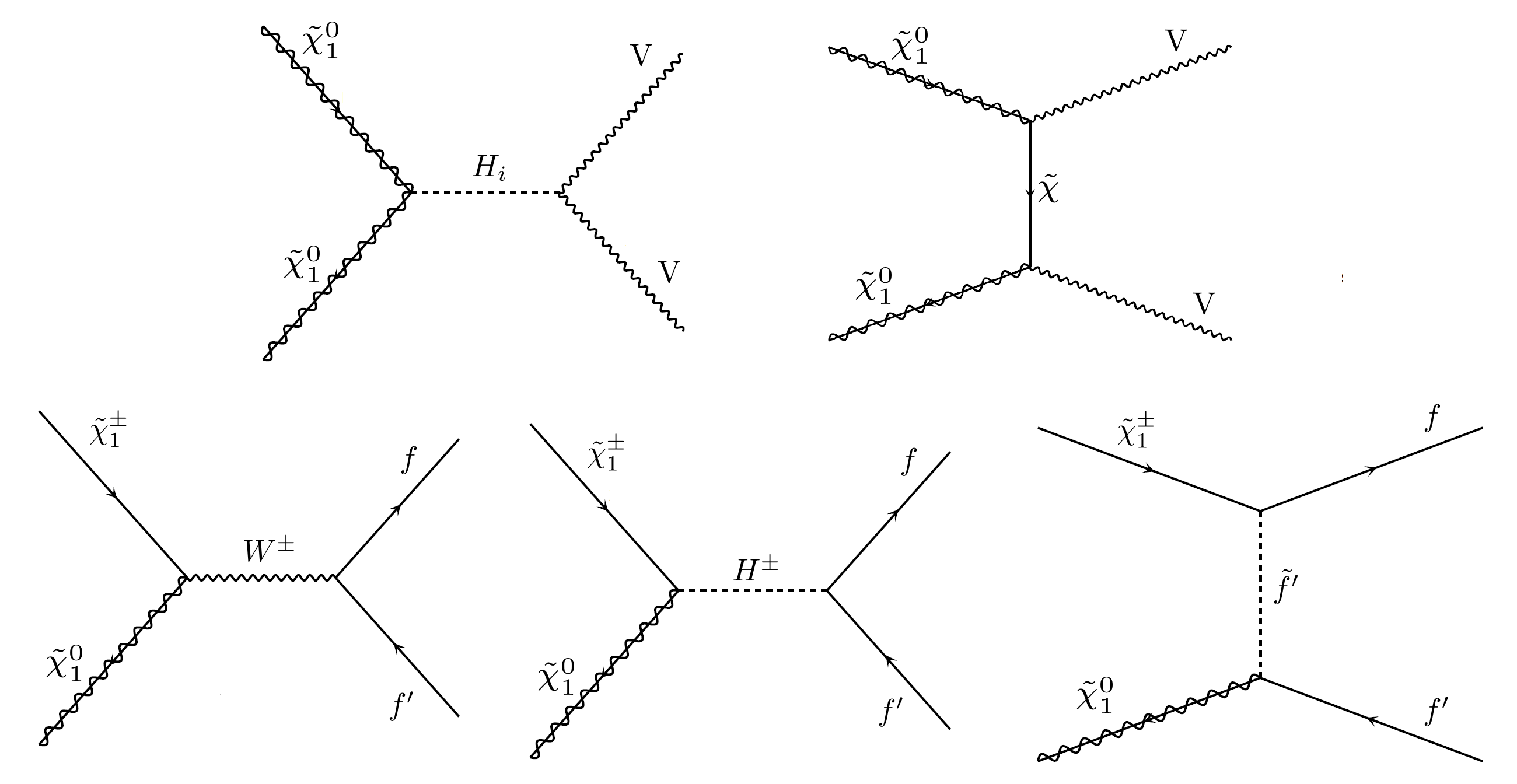}
	\end{center}
	\caption{
The Feynman diagrams for the main DM annihilation processes. In the first row, $V$ stands for $W$ boson or $Z$ boson, $H_i$ stands for the neutral Higgs bosons, $\tilde{\chi}$ stands for the charginos when $V=W$ and for the neutralinos when $V=Z$. Besides, the s-channel propagator can also be $Z$ boson when $V=W$. In the second row, $H^\pm$ is the charged Higgs bosons, 
$f,f^\prime$ are the SM fermions and $\tilde{f}^\prime$ are their SUSY partners.
}
	\label{feyn}
\end{figure}

We project surviving samples in Fig.~\ref{DM3}  on plots of $\mu^{H_1}_{VBF}+\mu^{H_2}_{VBF}$ - $C_{VBF,\tau\tau}$, $D_3$ - $C_{ttH,WW}$ and $\kappa$ - $C_{ggH,\gamma\gamma}$ in Fig.~\ref{DH6}.
It is clear that there is a strong correlation between $\sum_{i}\sigma_i^{H_1}/\sum_{i}\sigma_i^{h}$ and $C_{if}$. 
When the contribution on $\sum_{i}\sigma_i^{h}$ mainly comes from $H_1$ or $H_2$,  the values of 
$C_{if}$ are close to 1, as the samples with red or green shows. That is to say, these samples are consistent with the SM prediction. But there are lots of samples with $C_{if}$ smaller than 1, which means that the assumption with two mass-degenerate Higgs bosons may be true.
For example, in  Fig.~\ref{DH6}(a),  when $\mu_{VBF}^{H_1}+\mu_{VBF}^{H_2}$ is around the experimental measured value 
$\mu^{ex}_{VBF}=1.18$, samples have the lowest $C_{VBF,\tau\tau}$.
Fig.~\ref{DH6}(b) shows that $C_{ttH,WW}$ decreases obviously when $D_3$ is less than 1. 
In Fig.~\ref{DH6}(c), we can see that the SM prediction is better than samples in  the scenario with two mass-degenerate Higgs bosons
 for $C_{ggH,\gamma\gamma}$. 
The reason is that the cross section of $VBF$ in the scenario with two mass-degenerate Higgs bosons is raised compared to SM, but that of $ggH$ is not. 
When we take into account the total contribution $C_{t o t}\equiv \sum_{i, j} C_{i f}$
including the high precision processes	$(i,j)=(ggH,\gamma\gamma),(ggH,ZZ),(ggH,WW),(VBF,\gamma\gamma),(VBF,WW),(VBF,\tau\tau),(ttH,WWW)$ \cite{3,5,6}, there exist some samples with $C_{tot}<1$, which are more consistent with the measured Higgs data than the SM prediction.
We list the relevant parameters of two benchmark points in Table \ref{tab1}.
	For the DM mechanism that yields a relic density compatible with the observed upper bound, we find that
	the leading contribution is from the process $\tilde{\chi}_1^0 \tilde{\chi}_1^0 \to W^+ W^-$, 
	the contribution rate of this process is 49\% and 51\% for P1 and P2, respectively. 
	For P1, the subleading contribution with 27\% rate is from the coannihilation process $\tilde{\chi}_1^0 \tilde{\chi}_1^\pm \to f f^\prime$ due to the mass splitting between $\tilde{\chi}_1^\pm$ and $\tilde{\chi}_1^0$ at 11\%, 
	here $f,f^\prime$ are the SM quarks. 
	The other contribution is from the process $\tilde{\chi}_1^0 \tilde{\chi}_1^0 \to Z Z$. 
	For P2, the other contributions are from the processes $\tilde{\chi}_1^0 \tilde{\chi}_1^0 \to Z Z, H_i A_j, H_i H_j$,  
	here $H_i,A_i$ are the CP-even and CP-odd neutral Higgs, respectively. 
	The main annihilation Feynman diagrams for the two benchmark points are presented in Fig. \ref{feyn}.

\begin{table}[t]
	\centering
	\resizebox{1\textwidth}{!}
	{
		\begin{tabular}{crcr|crcr}
			\hline \hline
			\multicolumn{4}{c|}{\bf Benchmark Point P1}                                                                                                & \multicolumn{4}{c}{\bf Benchmark Point P2}                                                                                                \\ \hline
			$\lambda$             & 0.191     & $m_{\widetilde{\chi}_{1}^{0}}$   &105.2~GeV & $\lambda$             & 0.171 & $m_{\widetilde{\chi}_{1}^{0}}$       & 125.7~GeV \\
			$\kappa$              & 0.459     & $\left|N_{13}^{2}\right|$ & 0.479 & $\kappa$              & 0.318 & $\left|N_{13}^{2}\right|$ & 0.163 \\
			$\tan{\beta}$         & 17.47     & $\left|N_{14}^{2}\right|$& 0.455 & $\tan{\beta}$         & 13.41 & $\left|N_{14}^{2}\right|$ & 0.200 \\
			$\mu$                 & 72.49~GeV & $\left|N_{15}^{2}\right|$& 0.045 & $\mu$                 & 123.9~GeV & $\left|N_{15}^{2}\right|$& 0.611 \\
			$ \mu_{\rm tot}$ & 115.6~GeV & $\left|N_{13}^{2}-N_{14}^{2}\right|$ & 0.024 & $
			\mu_{\rm tot}$ & 161.8~GeV & $\left|N_{13}^{2}-N_{14}^{2}\right|$  & 0.037 \\
			$A_t$                 & -4218~GeV & 
			$\Omega h^2$   & 0.007 & 
			$A_t$                  & 2385~GeV & 
			$\Omega h^2$   & 0.043 \\
			$A_\kappa$            & -276.5~GeV  & $\sigma_{SD}[cm^2]$ & $1.8\times10^{-5}$ & $A_\kappa$                  & -77.21~GeV   & $\sigma_{SD}[pb]$   & $4.3\times10^{-5}$ \\
			$M_1$                 & 402.0~GeV          & $\sigma_{SI}[cm^2]$   & $1.2\times10^{-9}$ & 
			$M_1$                & 2327~GeV & 
			$\sigma_{SI}[pb]$   & $9.3\times10^{-11}$ \\
			$M_2$                 & 703.1~GeV          & $\mathcal{L}$   & 0.999 & 
			$M_2$            & 348.4~GeV & 
			$\mathcal{L}$    & 0.771 \\
			$m_{L_\mu}$                 & 226.4~GeV          & $D_B$   & 0.950 & 
			$m_{L_\mu}$                 & 335.7~GeV & 
			$D_B$   & 0.924 \\
			$m_E$                 & 933.5~GeV          & $C_{tot}$ & 0.922 & 
			$m_E$                 & 1373~GeV & 
			$C_{tot}$ & 0.965 
			
			\\ \hline \hline
			
	\end{tabular}}
	\caption{\label{tab1}
		Relevant parameters of the two benchmark points surviving all constraints from DM direct detections, the DM relic density, the Higgs data, sparticle searches at the LHC, and the $(g-2)_{\mu}$ measurement. $C_{tot}<1$ indicates that the scenario with two mass-degenerate Higgs bosons is more consistent with the measured data than the SM prediction.}
\end{table}

\subsection{$(g-2)_\mu$ in  the scenario with two mass-degenerate  SM-like Higgs bosons}
\begin{figure}[!htb]
	\begin{center}
		\includegraphics[width=0.5\textwidth]{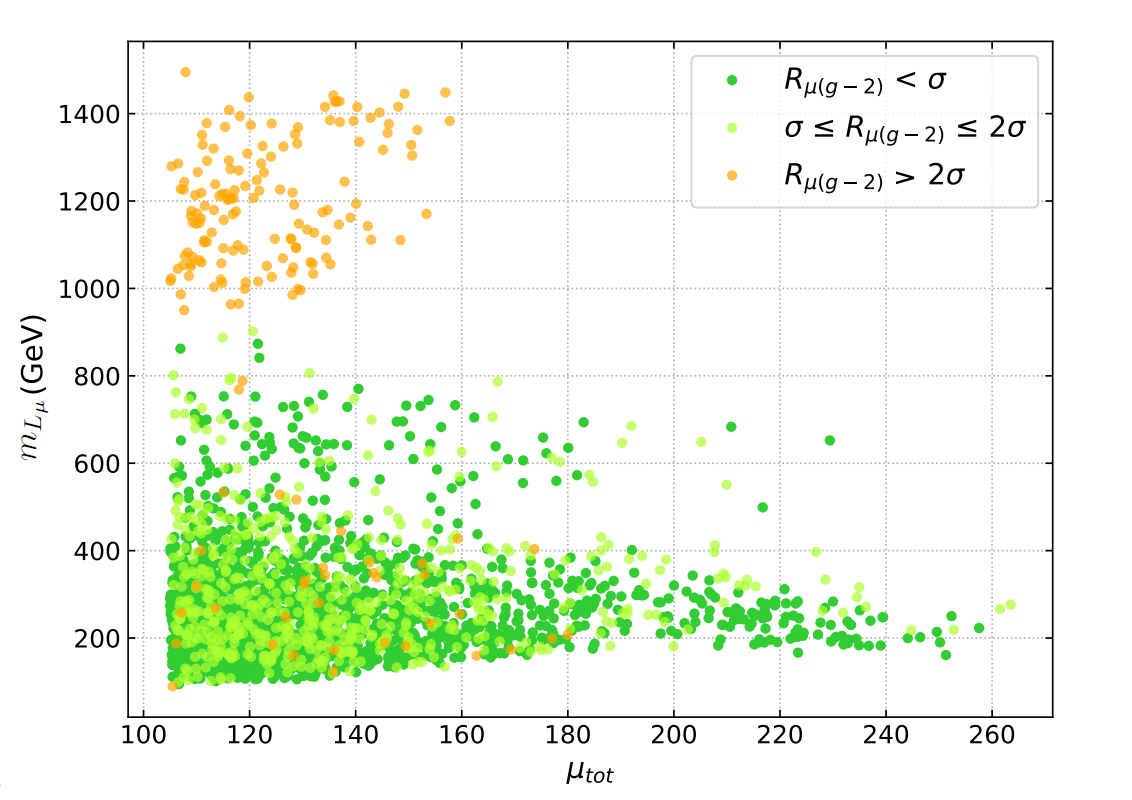}
	\end{center}
	\caption{
		All surviving samples in Fig.~\ref{DM3}  are projected on the plane $\mu_{tot}$ - $m_{L_{\mu}}$. The color of samples represents the degree of $(g-2)_\mu$ deviated from the experiment. Green samples represent deviations less than 1$\sigma$, light green samples represent deviations of 1$\sigma$ - 2$\sigma$, and yellow samples represent deviations more than 2$\sigma$.}
	\label{DH2}
\end{figure}

The muon anomalous magnetic $(g - 2)_\mu$  predicted by our collected samples are labeled as $a^{\text{SUSY}}_\mu$. We use SPheno-4.0.4 to calculate $(g-2)_\mu$, which only the BSM contributions are calculated (at the one-loop level), the SM-part is not calculated. The likelihood function $\mathcal{L}$ is defined as an indicator of the degree of conformity between $a^{\text{SUSY}}_\mu$ and the experiment\cite{17}:
\begin{equation}
	\label{L}
	\mathcal{L}=\exp \left[-\frac{1}{2}\left(\frac{a_{\mu}^{\mathrm{SUSY}}-2.51 \times 10^{-9}}{5.9 \times 10^{-10}}\right)^{2}\right]
\end{equation}
We project surviving samples on plot of $m_{L_{\mu}} - \mu_{tot}$ in Fig.~\ref{DH2}. 
We can see that it is difficult to obtain samples satisfying the result of $(g-2)_\mu$ experiment when $m_{L_{\mu}}$ is greater than 1000 GeV. 
This is because that the loop contribution to $(g-2)_\mu$ from smuon decreases sharply when smuon is much heavy.

\begin{figure}[!htb]
	\begin{center}
		\includegraphics[width=1\textwidth]{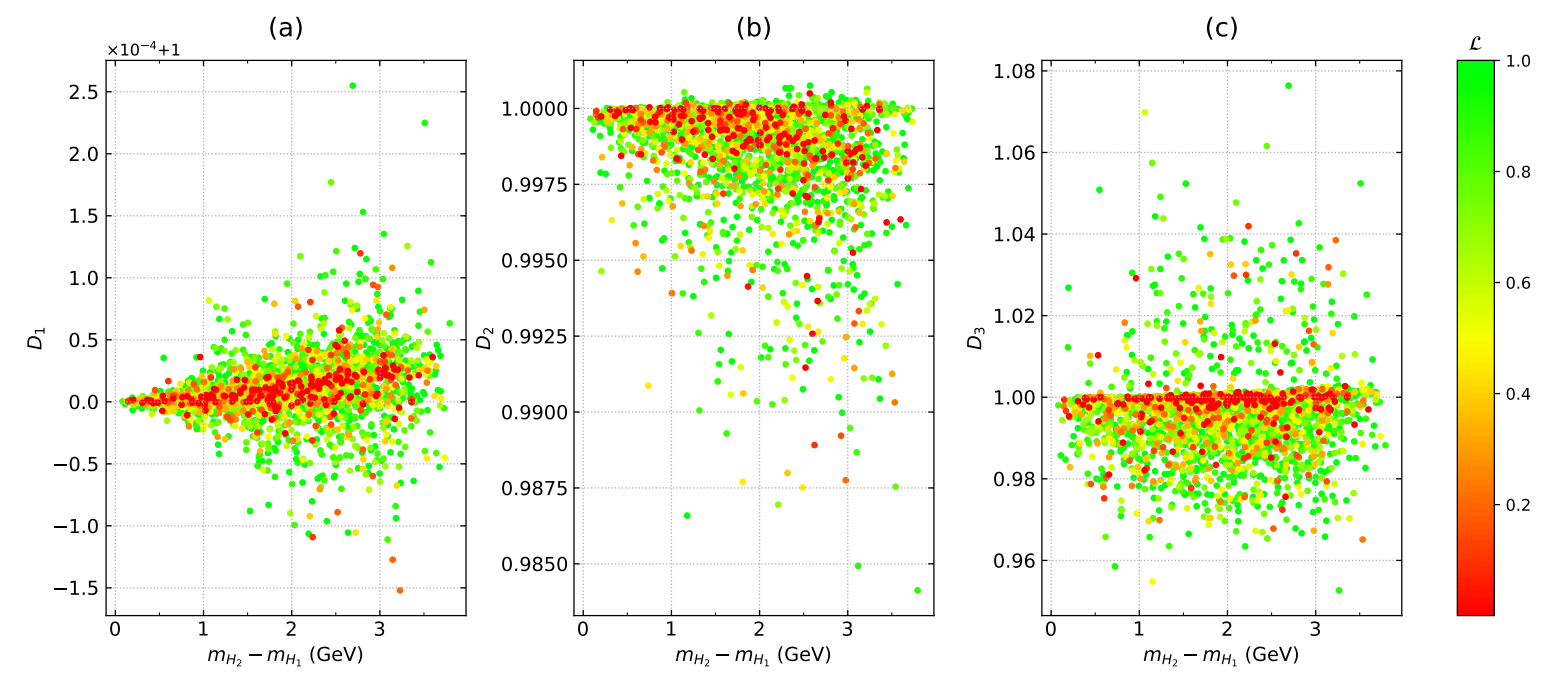}
	\end{center}
	\caption{All surviving samples in Fig.~\ref{DM3} are projected on plots $m_{H_2}-m_{H_1}$ versus $D_{i}(i=1,2,3)$. The colorbar represents the likelihood value in Eq.~(\ref{L}). Samples with green color are in better agreement with the recent experimental result of $(g-2)_{\mu}$, while samples with red color are in poorer agreement with the result.}
	\label{DH0}
\end{figure}

We project surviving samples on plots $m_{H_2}-m_{H_1}$ versus $D_{i}(i=1,2,3)$ in Fig.~\ref{DH0}.
We can see that the double ratios $D_{i}$ deviate from 1, indicating that there are different contributions from $H_1$ and $H_2$ on these ratios.
And there are lots of samples in agreement with the recent experimental result of $(g-2)_{\mu}$.
But we note that the deviations are not very outstanding. 
The reason is that while the production cross sections of two mass-degenerate Higgs bosons differ, the ratios are similar, i.e.,
$\sigma_{g g H}^{H_{1}} / \sigma_{V B F}^{H_{1}} \approx \sigma_{g g H}^{H_{2}} / \sigma_{V B F}^{H_{2}}$. 

To achieve better results, we can take an aggressive approach and only consider the decay branching ratios instead of the production cross-section:
\begin{equation}
	\label{DB}
	\begin{aligned}
		B_{Z Z, \gamma \gamma}&=\left(B_{Z Z}^{H_{1}} \times B_{\gamma \gamma}^{H_{1}}\right)+\left(B_{Z Z}^{H_{2}} \times B_{\gamma \gamma}^{H_{2}}\right), 
		B_{Z Z, W W}=\left(B_{Z Z}^{H_{1}} \times B_{W W}^{H_{1}}\right)+\left(B_{Z Z}^{H_{2}} \times B_{W W}^{H_{2}}\right) \\
		B_{b b, \gamma \gamma}=\left(B_{b b}^{H_{1}} \times B_{\gamma \gamma}^{H_{1}}\right) &+ \left(B_{b b}^{H_{2}} \times B_{\gamma \gamma}^{H_{2}}\right),
		B_{b b, W W}=\left(B_{b b}^{H_{1}} \times B_{W W}^{H_{1}}\right)+\left(B_{b b}^{H_{2}} \times B_{W W}^{H_{2}}\right),
		D_{B}=\frac{B_{Z Z, \gamma \gamma} / B_{Z Z, W W}}{B_{b b, \gamma \gamma} / B_{b b, W W}}
	\end{aligned}
\end{equation}
\begin{figure}[!htb]
	\begin{center}
		\includegraphics[width=0.5\textwidth]{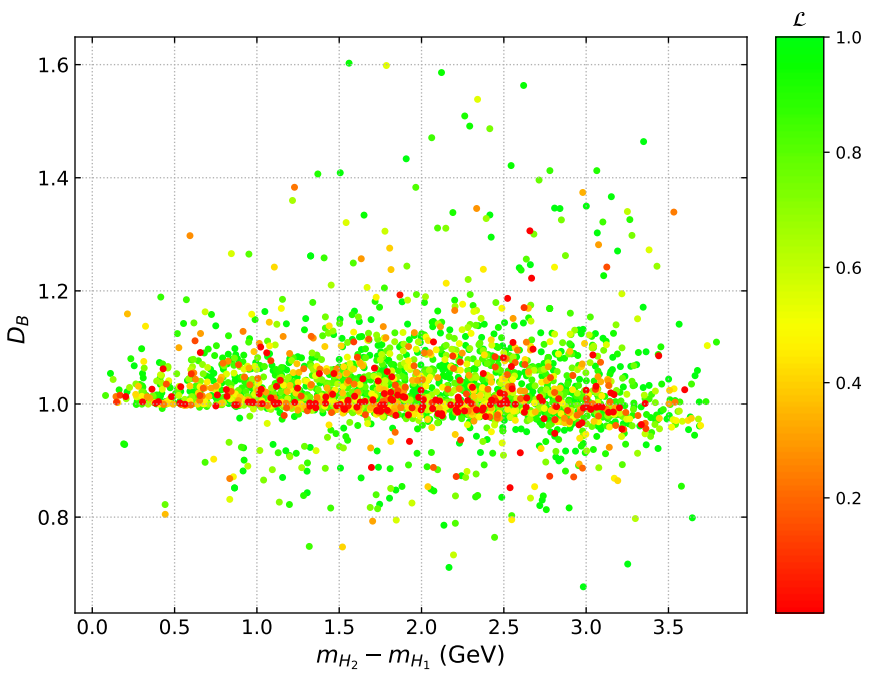}
	\end{center}
	\caption{
		Same as Fig.~\ref{DH0}, but samples are projected on plot $m_{H_2}-m_{H_1}$ versus $D_B$. }
	\label{DH1}
\end{figure}
$D_B$ is used to describe similarity of decay of $H_1$ and $H_2$. If $D_B$ is around 1, it means that decay features of $H_1$ and $H_2$ are the same. If $D_B$ deviates from 1, it means that the decay features are different.
We project surviving samples on plot $m_{H_2}-m_{H_1}$ versus $D_B$ in Fig.~\ref{DH1}.
We can see that the deviation is more obvious under the new variable $D_B$. 
There are some samples that typically deviate from 1. 
From the $(g-2)_\mu$ deviation distribution, we find that the recent experimental result from $(g-2)_\mu$ can be well explained in  the scenario with two mass-degenerate Higgs bosons. Samples with $D_B$ around 1 deviate from the center value of $(g-2)_\mu$, but some samples with $D_B$ deviation from 1 can satisfy the result of $(g-2)_\mu$ much better.

\section{\label{SU}Summary}
In the $\mu$NMSSM with two mass-degenerate SM-like Higgs bosons, we investigate the properties of DM, the observation of two mass-degenerate Higgs bosons and the explanation of $(g-2)_\mu$ anomaly.  We first scan the parameter space under current experimental constraints including results of DM direct detection experiments and from LHC searches for sparticles. We find that,
\begin{itemize}
	\item
	The SI and SD constraints are complementary in limiting the parameter space of  $\mu$NMSSM.  The DM direct detection limits favor the DM is higgsino-dominated.
        \item
	The scenario with two mass-degenerate SM-like Higgs bosons is capable of explaining the $(g-2)_\mu$ anomaly,
	that is, the scenarios we consider may be an
		explanation since they are compatible with the current measurements.
	\item 
	Deviations defined by the double ratios $D_i(i=1,2,3)$ are not very outstanding.
	The reason is that the ratio between the production cross section of $H_1$ and $H_2$ are similar. 
	To achieve better results, we define a new variable $D_B$. Samples with $D_B$ deviation from 1 can satisfy the result of $(g-2)_\mu$ much better.
	\item
	As far as the total contribution $C_{t o t}$ is concerned, there exist some samples more consistent with the measured Higgs data than the SM prediction.
			
\end{itemize}

\section{Acknowledgement}
We thank Junjie Cao, Bingfang Yang, Xinglong Jia and Shenshen Yang for helpful discussions. This work is supported by the National Natural Science Foundation of China (NNSFC) under grant No. 11705048, the National Research Project Cultivation Foundation of Henan Normal University under Grant Nos. 2021PL10. Besides, this work is powered by the High Performance Computing Center of Henan Normal University.

\appendix
\setcounter{table}{0}
\setcounter{figure}{0}
\renewcommand{\thetable}{A\arabic{table}}
\renewcommand{\thefigure}{A\arabic{figure}}

\section{Double ratios and components of Higgs boson in  the scenario with two mass-degenerate SM-like Higgs bosons}

\begin{figure}[!htb]
	\begin{center}
		\includegraphics[width=1\textwidth]{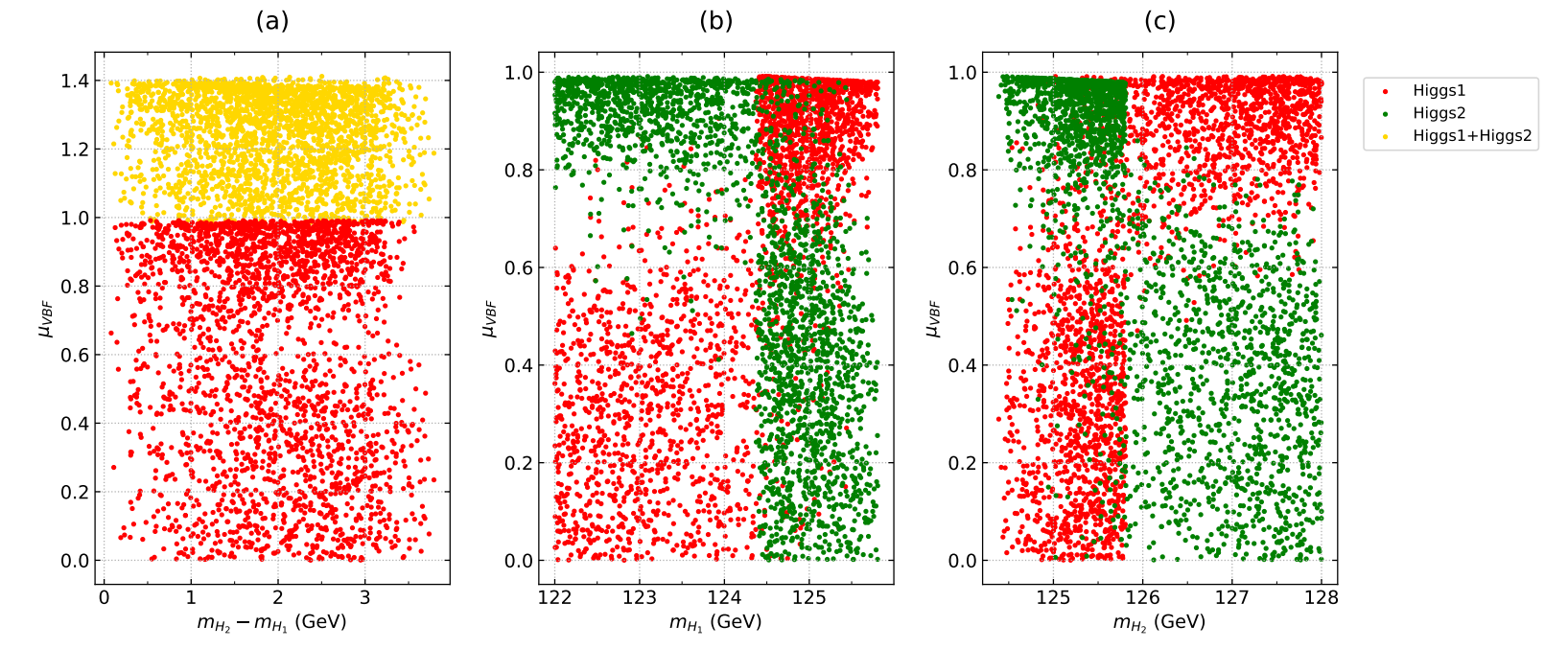}
	\end{center}
	\caption{
		All surviving samples in Fig.~\ref{DM3}  are projected on plots $m_{H_2}-m_{H_1}$ versus $\mu_{VBF}$, $m_{H_1}$ versus $\mu_{VBF}$, $m_{H_2}$ versus $\mu_{VBF}$. The red samples represent $\mu_{VBF}^{H_1} = \sigma_{VBF}^{H_1} / \sigma_{VBF}^{SM}$, the green samples represent $\mu_{VBF}^{H_2} = \sigma_{VBF}^{H_2} / \sigma_{VBF}^{SM}$, and the gold samples represent $\mu_{VBF}^{H_1}+\mu_{VBF}^{H_2}$.}
	\label{DH3}
\end{figure}

\begin{figure}[!htb]
	\begin{center}
		\includegraphics[width=0.9\textwidth]{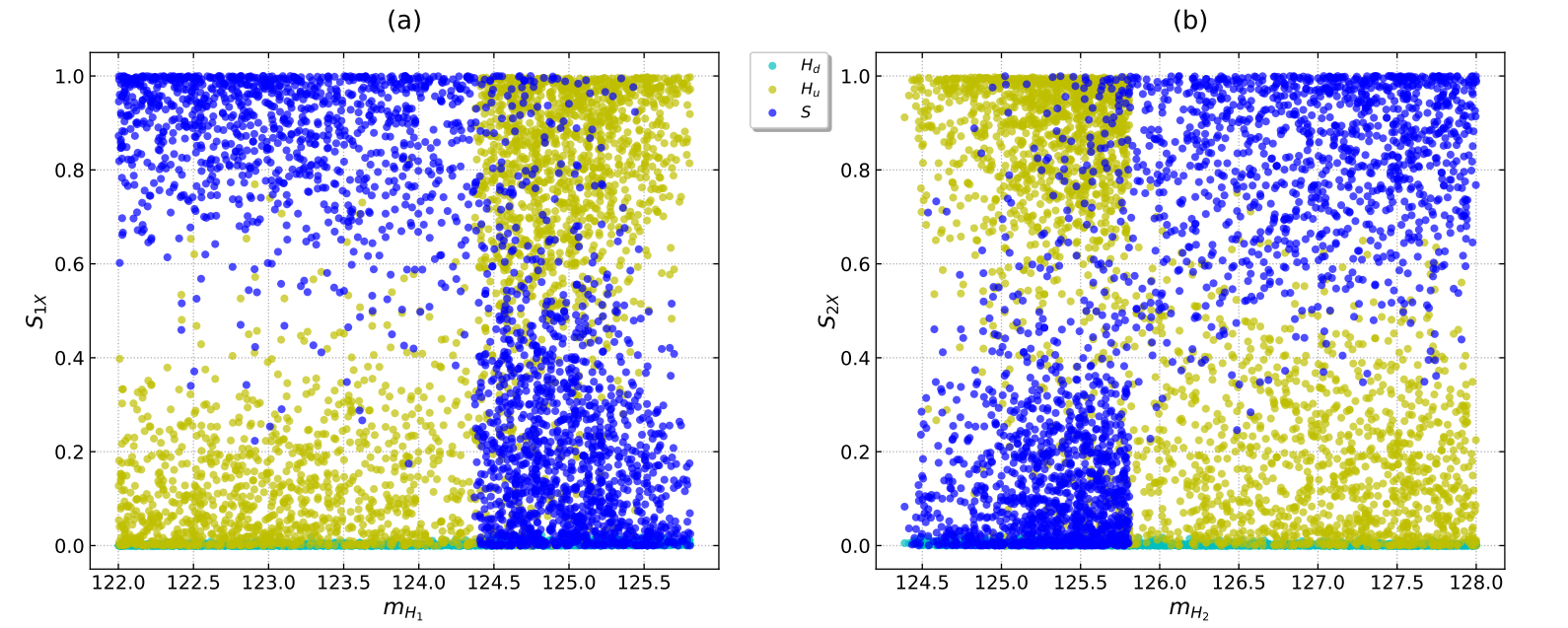}
	\end{center}
	\caption{
		All surviving samples in Fig.~\ref{DM3}  are projected on the planes $m_{H_1}$ - $S_{1X}$ and $m_{H_2}$ - $S_{2X}$. $H_1=S_{11} H_d+S_{12} H_u+S_{13} S$, $H_2=S_{21} H_d+S_{22} H_u+S_{23} S$,
		$S_{1X}$ and $S_{2X}$($X=1,2,3$) are proportions of each component $H_d$, $H_u$ and $S$. The light blue samples represent $S_{i1}(i=1,2)$, the yellow samples represent $S_{i2}$ and the blue samples represent $S_{i3}$.}
	\label{DH4}
\end{figure}

We project surviving samples on plots $m_{H_2}-m_{H_1}$ versus $\mu_{VBF}$, $m_{H_1}$ versus $\mu_{VBF}$, $m_{H_2}$ versus $\mu_{VBF}$
in Fig.~\ref{DH3}. We can see that the value of $\mu_{VBF}^{H_1}+\mu_{VBF}^{H_2}$ can be raised to the measured value 1.18 in Fig.~\ref{DH3},
but the single contribution $\mu_{VBF}^{H_1}$ or $\mu_{VBF}^{H_2}$ is all below 1.
We also project surviving samples on plots of $m_{H_1}$ - $S_{1X}$ and $m_{H_2}$ - $S_{2X}$ in Fig.~\ref{DH4}. 
Comparing Fig.~\ref{DH3}(b) with Fig.~\ref{DH4}(a) for $H_1$ and Fig.~\ref{DH3}(c) with Fig.~\ref{DH4}(b) for $H_2$, we can see that
when $H_1$ or $H_2$ are dominated by $H_u$, its contribution on $\sigma_{VBF}$ is large.
when $H_1$ or $H_2$ are dominated by $S$, its contribution on $\sigma_{VBF}$ becomes small. 
As a result, when the contribution on double ratios mainly comes from a single Higgs, the double ratios are around 1 according to Eq.~(\ref{D123}). 
Only if $H_1$ and $H_2$ are the mixing of $H_u$ and $S$, the double ratios can deviate from 1.

%


\begin{thebibliography}{99}
\vspace*{-1mm}
\begin{small}\baselineskip=10pt\itemsep-2pt
	
	
\bibitem{1}
B.~Abi \textit{et al.} [Muon g-2],
Measurement of the Positive Muon Anomalous Magnetic Moment to 0.46 ppm,
Phys. Rev. Lett. \textbf{126} (2021) no.14, 141801
[arXiv:2104.03281 [hep-ex]].

\bibitem{2}
G.~W.~Bennett \textit{et al.} [Muon g-2],
Final Report of the Muon E821 Anomalous Magnetic Moment Measurement at BNL,
Phys. Rev. D \textbf{73} (2006), 072003
[arXiv:hep-ex/0602035 [hep-ex]].

\bibitem{Aoyama:2020ynm}
T.~Aoyama, N.~Asmussen, M.~Benayoun, J.~Bijnens, T.~Blum, M.~Bruno, I.~Caprini, C.~M.~Carloni Calame, M.~C\`e and G.~Colangelo, \textit{et al.}
The anomalous magnetic moment of the muon in the Standard Model,
Phys. Rept. \textbf{887} (2020), 1-166
[arXiv:2006.04822 [hep-ph]].

\bibitem{c3}
Michel Davier, Andreas Hoecker, Bogdan Malaescu, and Zhiqing Zhang.
\newblock {Reevaluation of the hadronic vacuum polarisation contributions to
	the Standard Model predictions of the muon $g-2$ and ${\alpha (m_Z^2)}$ using
	newest hadronic cross-section data}.
\newblock {\em Eur. Phys. J.}, C77(12):827, 2017.

\bibitem{c4}
Alexander Keshavarzi, Daisuke Nomura, and Thomas Teubner.
\newblock {Muon $g-2$ and $\alpha(M_Z^2)$: a new data-based analysis}.
\newblock {\em Phys. Rev.}, D97(11):114025, 2018.

\bibitem{c5}
Gilberto Colangelo, Martin Hoferichter, and Peter Stoffer.
\newblock {Two-pion contribution to hadronic vacuum polarization}.
\newblock {\em JHEP}, 02:006, 2019.

\bibitem{c6}
Martin Hoferichter, Bai-Long Hoid, and Bastian Kubis.
\newblock {Three-pion contribution to hadronic vacuum polarization}.
\newblock {\em JHEP}, 08:137, 2019.

\bibitem{c7}
M.~Davier, A.~Hoecker, B.~Malaescu, and Z.~Zhang.
\newblock {A new evaluation of the hadronic vacuum polarisation contributions
	to the muon anomalous magnetic moment and to
	$\mathbf{\boldsymbol\alpha(m_Z^2)}$}.
\newblock {\em Eur. Phys. J.}, C80(3):241, 2020.
\newblock [Erratum: Eur. Phys. J. {\bf C80}, 410 (2020)].

\bibitem{c8}
Alexander Keshavarzi, Daisuke Nomura, and Thomas Teubner.
\newblock {The $g-2$ of charged leptons, $\alpha(M_Z^2)$ and the hyperfine
	splitting of muonium}.
\newblock {\em Phys. Rev.}, D101:014029, 2020.

\bibitem{c9}
Alexander Kurz, Tao Liu, Peter Marquard, and Matthias Steinhauser.
\newblock {Hadronic contribution to the muon anomalous magnetic moment to
	next-to-next-to-leading order}.
\newblock {\em Phys. Lett.}, B734:144--147, 2014.

\bibitem{c10}
B.~Chakraborty et~al.
\newblock {Strong-Isospin-Breaking Correction to the Muon Anomalous Magnetic
	Moment from Lattice QCD at the Physical Point}.
\newblock {\em Phys. Rev. Lett.}, 120(15):152001, 2018.

\bibitem{c11}
Sz. Borsanyi et~al.
\newblock {Hadronic vacuum polarization contribution to the anomalous magnetic
	moments of leptons from first principles}.
\newblock {\em Phys. Rev. Lett.}, 121(2):022002, 2018.

\bibitem{c12}
T.~Blum, P.~A. Boyle, V.~G{\"u}lpers, T.~Izubuchi, L.~Jin, C.~Jung,
A.~J{\"u}ttner, C.~Lehner, A.~Portelli, and J.~T. Tsang.
\newblock {Calculation of the hadronic vacuum polarization contribution to the
	muon anomalous magnetic moment}.
\newblock {\em Phys. Rev. Lett.}, 121(2):022003, 2018.

\bibitem{c13}
D.~Giusti, V.~Lubicz, G.~Martinelli, F.~Sanfilippo, and S.~Simula.
\newblock {Electromagnetic and strong isospin-breaking corrections to the muon
	$g - 2$ from Lattice QCD+QED}.
\newblock {\em Phys. Rev.}, D99(11):114502, 2019.

\bibitem{c14}
Eigo Shintani and Yoshinobu Kuramashi.
\newblock {Study of systematic uncertainties in hadronic vacuum polarization
	contribution to muon $g-2$ with 2+1 flavor lattice QCD}.
\newblock {\em Phys. Rev.}, D100(3):034517, 2019.

\bibitem{c15}
C.~T.~H. Davies et~al.
\newblock {Hadronic-vacuum-polarization contribution to the muon's anomalous
	magnetic moment from four-flavor lattice QCD}.
\newblock {\em Phys. Rev.}, D101(3):034512, 2020.

\bibitem{c16}
Antoine G\'erardin, Marco C\`e, Georg von Hippel, Ben H{\"o}rz, Harvey~B.
Meyer, Daniel Mohler, Konstantin Ottnad, Jonas Wilhelm, and Hartmut Wittig.
\newblock {The leading hadronic contribution to $(g-2)_\mu$ from lattice QCD
	with $N_{\rm f}=2+1$ flavours of O($a$) improved Wilson quarks}.
\newblock {\em Phys. Rev.}, D100(1):014510, 2019.

\bibitem{c17}
Christopher Aubin, Thomas Blum, Cheng Tu, Maarten Golterman, Chulwoo Jung, and
Santiago Peris.
\newblock {Light quark vacuum polarization at the physical point and
	contribution to the muon $g-2$}.
\newblock {\em Phys. Rev.}, D101(1):014503, 2020.

\bibitem{c18}
D.~Giusti and S.~Simula.
\newblock {Lepton anomalous magnetic moments in Lattice QCD+QED}.
\newblock {\em PoS}, LATTICE2019:104, 2019.

\bibitem{c19}
Kirill Melnikov and Arkady Vainshtein.
\newblock {Hadronic light-by-light scattering contribution to the muon
	anomalous magnetic moment revisited}.
\newblock {\em Phys. Rev.}, D70:113006, 2004.

\bibitem{c20}
Pere Masjuan and Pablo S{\'a}nchez-Puertas.
\newblock {Pseudoscalar-pole contribution to the $(g_{\mu}-2)$: a rational
	approach}.
\newblock {\em Phys. Rev.}, D95(5):054026, 2017.

\bibitem{c21}
Gilberto Colangelo, Martin Hoferichter, Massimiliano Procura, and Peter
Stoffer.
\newblock {Dispersion relation for hadronic light-by-light scattering: two-pion
	contributions}.
\newblock {\em JHEP}, 04:161, 2017.

\bibitem{c22}
Martin Hoferichter, Bai-Long Hoid, Bastian Kubis, Stefan Leupold, and
Sebastian~P. Schneider.
\newblock {Dispersion relation for hadronic light-by-light scattering: pion
	pole}.
\newblock {\em JHEP}, 10:141, 2018.

\bibitem{c23}
Antoine G{\'e}rardin, Harvey~B. Meyer, and Andreas Nyffeler.
\newblock {Lattice calculation of the pion transition form factor with
	$N_f=2+1$ Wilson quarks}.
\newblock {\em Phys. Rev.}, D100(3):034520, 2019.

\bibitem{c24}
Johan Bijnens, Nils Hermansson-Truedsson, and Antonio
Rodr{\'i}guez-S{\'a}nchez.
\newblock {Short-distance constraints for the HLbL contribution to the muon
	anomalous magnetic moment}.
\newblock {\em Phys. Lett.}, B798:134994, 2019.

\bibitem{c25}
Gilberto Colangelo, Franziska Hagelstein, Martin Hoferichter, Laetitia Laub,
and Peter Stoffer.
\newblock {Longitudinal short-distance constraints for the hadronic
	light-by-light contribution to $(g-2)_\mu$ with large-$N_c$ Regge models}.
\newblock {\em JHEP}, 03:101, 2020.

\bibitem{c26}
Vladyslav Pauk and Marc Vanderhaeghen.
\newblock {Single meson contributions to the muon`s anomalous magnetic moment}.
\newblock {\em Eur. Phys. J.}, C74(8):3008, 2014.

\bibitem{c27}
Igor Danilkin and Marc Vanderhaeghen.
\newblock {Light-by-light scattering sum rules in light of new data}.
\newblock {\em Phys. Rev.}, D95(1):014019, 2017.

\bibitem{c28}
Friedrich Jegerlehner.
\newblock {The Anomalous Magnetic Moment of the Muon}.
\newblock {\em Springer Tracts Mod. Phys.}, 274:1--693, 2017.

\bibitem{c29}
M.~Knecht, S.~Narison, A.~Rabemananjara, and D.~Rabetiarivony.
\newblock {Scalar meson contributions to $a_\mu$ from hadronic light-by-light
	scattering}.
\newblock {\em Phys. Lett.}, B787:111--123, 2018.

\bibitem{c30}
Gernot Eichmann, Christian~S. Fischer, and Richard Williams.
\newblock {Kaon-box contribution to the anomalous magnetic moment of the muon}.
\newblock {\em Phys. Rev.}, D101(5):054015, 2020.

\bibitem{c31}
Pablo Roig and Pablo S{\'a}nchez-Puertas.
\newblock {Axial-vector exchange contribution to the hadronic light-by-light
	piece of the muon anomalous magnetic moment}.
\newblock {\em Phys. Rev.}, D101(7):074019, 2020.

\bibitem{c32}
Gilberto Colangelo, Martin Hoferichter, Andreas Nyffeler, Massimo Passera, and
Peter Stoffer.
\newblock {Remarks on higher-order hadronic corrections to the muon $g-2$}.
\newblock {\em Phys. Lett.}, B735:90--91, 2014.

\bibitem{c33}
Thomas Blum, Norman Christ, Masashi Hayakawa, Taku Izubuchi, Luchang Jin,
Chulwoo Jung, and Christoph Lehner.
\newblock {The hadronic light-by-light scattering contribution to the muon
	anomalous magnetic moment from lattice QCD}.
\newblock {\em Phys. Rev. Lett.}, 124(13):132002, 2020.

\bibitem{c34}
Tatsumi Aoyama, Masashi Hayakawa, Toichiro Kinoshita, and Makiko Nio.
\newblock {Complete Tenth-Order QED Contribution to the Muon $g-2$}.
\newblock {\em Phys. Rev. Lett.}, 109:111808, 2012.

\bibitem{c35}
Tatsumi Aoyama, Toichiro Kinoshita, and Makiko Nio.
\newblock {Theory of the Anomalous Magnetic Moment of the Electron}.
\newblock {\em Atoms}, 7(1):28, 2019.

\bibitem{c36}
Andrzej Czarnecki, William~J. Marciano, and Arkady Vainshtein.
\newblock {Refinements in electroweak contributions to the muon anomalous
	magnetic moment}.
\newblock {\em Phys. Rev.}, D67:073006, 2003.
\newblock [Erratum: Phys. Rev. {\bf D73}, 119901 (2006)].

\bibitem{c37}
C.~Gnendiger, D.~St{\"o}ckinger, and H.~St{\"o}ckinger-Kim.
\newblock {The electroweak contributions to $(g-2)_\mu$ after the Higgs boson
	mass measurement}.
\newblock {\em Phys. Rev.}, D88:053005, 2013.

\bibitem{3a0}
P.~Athron, C.~Bal\'azs, D.~H.~J.~Jacob, W.~Kotlarski, D.~St\"ockinger and H.~St\"ockinger-Kim,
New physics explanations of a$_{\mu}$ in light of the FNAL muon $g-2$ measurement,
JHEP \textbf{09}, 080 (2021)
doi:10.1007/JHEP09(2021)080
[arXiv:2104.03691 [hep-ph]].

\bibitem{3a1}
S.~P.~Martin and J.~D.~Wells,
Muon Anomalous Magnetic Dipole Moment in Supersymmetric Theories,
Phys. Rev. D \textbf{64} (2001), 035003
[arXiv:hep-ph/0103067 [hep-ph]].

\bibitem{3a2}
A.~Czarnecki and W.~J.~Marciano,
The Muon anomalous magnetic moment: A Harbinger for 'new physics',
Phys. Rev. D \textbf{64} (2001), 013014
[arXiv:hep-ph/0102122 [hep-ph]].

\bibitem{3a3}
D.~Stockinger,
The Muon Magnetic Moment and Supersymmetry,
J. Phys. G \textbf{34} (2007), R45-R92
[arXiv:hep-ph/0609168 [hep-ph]].

\bibitem{li1}
P.~Athron, M.~Bach, H.~G.~Fargnoli, C.~Gnendiger, R.~Greifenhagen, J.~h.~Park, S.~Pa\ss{}ehr, D.~St\"ockinger, H.~St\"ockinger-Kim and A.~Voigt,
GM2Calc: Precise MSSM prediction for $(g - 2)$ of the muon,
Eur. Phys. J. C \textbf{76}, no.2, 62 (2016)
doi:10.1140/epjc/s10052-015-3870-2
[arXiv:1510.08071 [hep-ph]].

\bibitem{li4}
C.~H.~Chen and C.~Q.~Geng,
The Muon anomalous magnetic moment from a generic charged Higgs with SUSY,
Phys. Lett. B \textbf{511}, 77-84 (2001)
doi:10.1016/S0370-2693(01)00651-7
[arXiv:hep-ph/0104151 [hep-ph]].
\bibitem{li5}
A.~Arhrib and S.~Baek,
Two loop Barr-Zee type contributions to (g-2)(muon) in the MSSM,
Phys. Rev. D \textbf{65}, 075002 (2002)
doi:10.1103/PhysRevD.65.075002
[arXiv:hep-ph/0104225 [hep-ph]].
\bibitem{li6}
M.~Byrne, C.~Kolda and J.~E.~Lennon,
Updated implications of the muon anomalous magnetic moment for supersymmetry,
Phys. Rev. D \textbf{67}, 075004 (2003)
doi:10.1103/PhysRevD.67.075004
[arXiv:hep-ph/0208067 [hep-ph]].
\bibitem{li7}
K.~Cheung, O.~C.~W.~Kong and J.~S.~Lee,
Electric and anomalous magnetic dipole moments of the muon in the MSSM,
JHEP \textbf{06}, 020 (2009)
doi:10.1088/1126-6708/2009/06/020
[arXiv:0904.4352 [hep-ph]].
\bibitem{li8}
S.~Heinemeyer, D.~Stockinger and G.~Weiglein,
Two loop SUSY corrections to the anomalous magnetic moment of the muon,
Nucl. Phys. B \textbf{690}, 62-80 (2004)
doi:10.1016/j.nuclphysb.2004.04.017
[arXiv:hep-ph/0312264 [hep-ph]].
\bibitem{li9}
S.~Heinemeyer, D.~Stockinger and G.~Weiglein,
Electroweak and supersymmetric two-loop corrections to (g-2)(mu),
Nucl. Phys. B \textbf{699}, 103-123 (2004)
doi:10.1016/j.nuclphysb.2004.08.014
[arXiv:hep-ph/0405255 [hep-ph]].
\bibitem{li10}
T.~F.~Feng, X.~Q.~Li, L.~Lin, J.~Maalampi and H.~S.~Song,
The Two-loop supersymmetric corrections to lepton anomalous magnetic and electric dipole moments,
Phys. Rev. D \textbf{73}, 116001 (2006)
doi:10.1103/PhysRevD.73.116001
[arXiv:hep-ph/0604171 [hep-ph]].
\bibitem{li11}
S.~Marchetti, S.~Mertens, U.~Nierste and D.~Stockinger,
Tan(beta)-enhanced supersymmetric corrections to the anomalous magnetic moment of the muon,
Phys. Rev. D \textbf{79}, 013010 (2009)
doi:10.1103/PhysRevD.79.013010
[arXiv:0808.1530 [hep-ph]].
\bibitem{li12}
P.~von Weitershausen, M.~Schafer, H.~Stockinger-Kim and D.~Stockinger,
Photonic SUSY Two-Loop Corrections to the Muon Magnetic Moment,
Phys. Rev. D \textbf{81}, 093004 (2010)
doi:10.1103/PhysRevD.81.093004
[arXiv:1003.5820 [hep-ph]].
\bibitem{li13}
H.~G.~Fargnoli, C.~Gnendiger, S.~Pa\ss{}ehr, D.~St\"ockinger and H.~St\"ockinger-Kim,
Non-decoupling two-loop corrections to $(g-2)_\mu$ from fermion/sfermion loops in the MSSM,
Phys. Lett. B \textbf{726}, 717-724 (2013)
doi:10.1016/j.physletb.2013.09.034
[arXiv:1309.0980 [hep-ph]].
\bibitem{li14}
H.~Fargnoli, C.~Gnendiger, S.~Pa\ss{}ehr, D.~St\"ockinger and H.~St\"ockinger-Kim,
Two-loop corrections to the muon magnetic moment from fermion/sfermion loops in the MSSM: detailed results,
JHEP \textbf{02}, 070 (2014)
doi:10.1007/JHEP02(2014)070
[arXiv:1311.1775 [hep-ph]].
\bibitem{li15}
G.~C.~Cho, K.~Hagiwara, Y.~Matsumoto and D.~Nomura,
The MSSM confronts the precision electroweak data and the muon g-2,
JHEP \textbf{11}, 068 (2011)
doi:10.1007/JHEP11(2011)068
[arXiv:1104.1769 [hep-ph]].

\bibitem{li17}
M.~Endo, K.~Hamaguchi, S.~Iwamoto, K.~Nakayama and N.~Yokozaki,
Higgs mass and muon anomalous magnetic moment in the U(1) extended MSSM,
Phys. Rev. D \textbf{85}, 095006 (2012)
doi:10.1103/PhysRevD.85.095006
[arXiv:1112.6412 [hep-ph]].
\bibitem{li18}
M.~Endo, K.~Hamaguchi, S.~Iwamoto and N.~Yokozaki,
Higgs Mass and Muon Anomalous Magnetic Moment in Supersymmetric Models with Vector-Like Matters,
Phys. Rev. D \textbf{84}, 075017 (2011)
doi:10.1103/PhysRevD.84.075017
[arXiv:1108.3071 [hep-ph]].
\bibitem{li19}
M.~Ibe, S.~Matsumoto, T.~T.~Yanagida and N.~Yokozaki,
Heavy Squarks and Light Sleptons in Gauge Mediation \textasciitilde{}From the viewpoint of 125 GeV Higgs Boson and Muon g-2\textasciitilde{},
JHEP \textbf{03}, 078 (2013)
doi:10.1007/JHEP03(2013)078
[arXiv:1210.3122 [hep-ph]].
\bibitem{li20}
J.~E.~Kim,
Inverted effective supersymmetry with combined Z' and gravity mediation, and muon anomalous magnetic moment,
Phys. Rev. D \textbf{87}, no.1, 015004 (2013)
doi:10.1103/PhysRevD.87.015004
[arXiv:1208.5484 [hep-ph]].
\bibitem{li21}
M.~Endo, K.~Hamaguchi, S.~Iwamoto and T.~Yoshinaga,
Muon g-2 vs LHC in Supersymmetric Models,
JHEP \textbf{01}, 123 (2014)
doi:10.1007/JHEP01(2014)123
[arXiv:1303.4256 [hep-ph]].
\bibitem{li22}
M.~Ibe, T.~T.~Yanagida and N.~Yokozaki,
Muon g-2 and 125 GeV Higgs in Split-Family Supersymmetry,
JHEP \textbf{08}, 067 (2013)
doi:10.1007/JHEP08(2013)067
[arXiv:1303.6995 [hep-ph]].
\bibitem{li23}
S.~Akula and P.~Nath,
Gluino-driven radiative breaking, Higgs boson mass, muon g-2, and the Higgs diphoton decay in supergravity unification,
Phys. Rev. D \textbf{87}, no.11, 115022 (2013)
doi:10.1103/PhysRevD.87.115022
[arXiv:1304.5526 [hep-ph]].
\bibitem{li24}
H.~B.~Zhang, T.~F.~Feng, S.~M.~Zhao and T.~J.~Gao,
Lepton-flavor violation and $(g-2)_\mu$ in the $\mu\nu$SSM,
Nucl. Phys. B \textbf{873}, 300-324 (2013)
[erratum: Nucl. Phys. B \textbf{879}, 235 (2014)]
doi:10.1016/j.nuclphysb.2013.04.018
[arXiv:1304.6248 [hep-ph]].
\bibitem{li25}
M.~Endo, K.~Hamaguchi, T.~Kitahara and T.~Yoshinaga,
Probing Bino contribution to muon $g - 2$,
JHEP \textbf{11}, 013 (2013)
doi:10.1007/JHEP11(2013)013
[arXiv:1309.3065 [hep-ph]].
\bibitem{li26}
G.~Bhattacharyya, B.~Bhattacherjee, T.~T.~Yanagida and N.~Yokozaki,
A practical GMSB model for explaining the muon (g-2) with gauge coupling unification,
Phys. Lett. B \textbf{730}, 231-235 (2014)
doi:10.1016/j.physletb.2013.12.064
[arXiv:1311.1906 [hep-ph]].
\bibitem{li27}
J.~L.~Evans, M.~Ibe, K.~A.~Olive and T.~T.~Yanagida,
One-loop anomaly mediated scalar masses and $(g-2)_mu$ in pure gravity mediation,
Eur. Phys. J. C \textbf{74}, no.2, 2775 (2014)
doi:10.1140/epjc/s10052-014-2775-9
[arXiv:1312.1984 [hep-ph]].
\bibitem{li28}
S.~Iwamoto, T.~T.~Yanagida and N.~Yokozaki,
CP-safe gravity mediation and muon g \ensuremath{-} 2,
PTEP \textbf{2015}, 073B01 (2015)
doi:10.1093/ptep/ptv084
[arXiv:1407.4226 [hep-ph]].
\bibitem{li29}
J.~Kersten, J.~h.~Park, D.~St\"ockinger and L.~Velasco-Sevilla,
Understanding the correlation between $(g-2)_\mu$ and $\mu \rightarrow e \gamma$ in the MSSM,
JHEP \textbf{08}, 118 (2014)
doi:10.1007/JHEP08(2014)118
[arXiv:1405.2972 [hep-ph]].
\bibitem{li30}
I.~Gogoladze, F.~Nasir, Q.~Shafi and C.~S.~Un,
Nonuniversal Gaugino Masses and Muon g-2,
Phys. Rev. D \textbf{90}, no.3, 035008 (2014)
doi:10.1103/PhysRevD.90.035008
[arXiv:1403.2337 [hep-ph]].
\bibitem{li31}
M.~Badziak, Z.~Lalak, M.~Lewicki, M.~Olechowski and S.~Pokorski,
Upper bounds on sparticle masses from muon g \ensuremath{-} 2 and the Higgs mass and the complementarity of future colliders,
JHEP \textbf{03}, 003 (2015)
doi:10.1007/JHEP03(2015)003
[arXiv:1411.1450 [hep-ph]].
\bibitem{li32}
K.~Kowalska, L.~Roszkowski, E.~M.~Sessolo and A.~J.~Williams,
GUT-inspired SUSY and the muon g \ensuremath{-} 2 anomaly: prospects for LHC 14 TeV,
JHEP \textbf{06}, 020 (2015)
doi:10.1007/JHEP06(2015)020
[arXiv:1503.08219 [hep-ph]].
\bibitem{li33}
J.~Chakrabortty, A.~Choudhury and S.~Mondal,
Non-universal Gaugino mass models under the lamppost of muon (g-2),
JHEP \textbf{07}, 038 (2015)
doi:10.1007/JHEP07(2015)038
[arXiv:1503.08703 [hep-ph]].
\bibitem{li34}
B.~P.~Padley, K.~Sinha and K.~Wang,
Natural Supersymmetry, Muon $g-2$, and the Last Crevices for the Top Squark,
Phys. Rev. D \textbf{92}, no.5, 055025 (2015)
doi:10.1103/PhysRevD.92.055025
[arXiv:1505.05877 [hep-ph]].
\bibitem{li35}
M.~Bach, J.~h.~Park, D.~St\"ockinger and H.~St\"ockinger-Kim,
Large muon $(g-2)$ with TeV-scale SUSY masses for $\tan\beta\to\infty$,
JHEP \textbf{10}, 026 (2015)
doi:10.1007/JHEP10(2015)026
[arXiv:1504.05500 [hep-ph]].
\bibitem{li36}
K.~Harigaya, T.~T.~Yanagida and N.~Yokozaki,
Muon g\ensuremath{-}2 in focus point SUSY,
Phys. Rev. D \textbf{92}, no.3, 035011 (2015)
doi:10.1103/PhysRevD.92.035011
[arXiv:1505.01987 [hep-ph]].
\bibitem{li37}
D.~Chowdhury and N.~Yokozaki,
Muon g \ensuremath{-} 2 in anomaly mediated SUSY breaking,
JHEP \textbf{08}, 111 (2015)
doi:10.1007/JHEP08(2015)111
[arXiv:1505.05153 [hep-ph]].
\bibitem{li38}
S.~Khalil and C.~S.~Un,
Muon Anomalous Magnetic Moment in SUSY B-L Model with Inverse Seesaw,
Phys. Lett. B \textbf{763}, 164-168 (2016)
doi:10.1016/j.physletb.2016.10.035
[arXiv:1509.05391 [hep-ph]].
\bibitem{li39}
M.~A.~Ajaib, B.~Dutta, T.~Ghosh, I.~Gogoladze and Q.~Shafi,
Neutralinos and sleptons at the LHC in light of muon $(g-2)_{\mu}$,
Phys. Rev. D \textbf{92}, no.7, 075033 (2015)
doi:10.1103/PhysRevD.92.075033
[arXiv:1505.05896 [hep-ph]].
\bibitem{li40}
K.~Harigaya, T.~T.~Yanagida and N.~Yokozaki,
Higgs boson mass of 125 GeV and $g-2$ of the muon in a gaugino mediation model,
Phys. Rev. D \textbf{91}, no.7, 075010 (2015)
doi:10.1103/PhysRevD.91.075010
[arXiv:1501.07447 [hep-ph]].
\bibitem{li41}
I.~Gogoladze, Q.~Shafi and C.~S.~\"Un,
Reconciling the muon g\ensuremath{-}2 , a 125 GeV Higgs boson, and dark matter in gauge mediation models,
Phys. Rev. D \textbf{92}, no.11, 115014 (2015)
doi:10.1103/PhysRevD.92.115014
[arXiv:1509.07906 [hep-ph]].
\bibitem{li42}
F.~V.~Flores-Baez, M.~G\'omez Bock and M.~Mondrag\'on,
Muon g-2 through a flavor structure on soft SUSY terms,
Eur. Phys. J. C \textbf{76}, no.10, 561 (2016)
doi:10.1140/epjc/s10052-016-4402-4
[arXiv:1512.00902 [hep-ph]].
\bibitem{li43}
A.~S.~Belyaev, J.~E.~Camargo-Molina, S.~F.~King, D.~J.~Miller, A.~P.~Morais and P.~B.~Schaefers,
A to Z of the Muon Anomalous Magnetic Moment in the MSSM with Pati-Salam at the GUT scale,
JHEP \textbf{06}, 142 (2016)
doi:10.1007/JHEP06(2016)142
[arXiv:1605.02072 [hep-ph]].
\bibitem{li44}
T.~Li, S.~Raza and K.~Wang,
Constraining Natural SUSY via the Higgs Coupling and the Muon Anomalous Magnetic Moment Measurements,
Phys. Rev. D \textbf{93}, no.5, 055040 (2016)
doi:10.1103/PhysRevD.93.055040
[arXiv:1601.00178 [hep-ph]].
\bibitem{li45}
N.~Okada and H.~M.~Tran,
125 GeV Higgs boson mass and muon $g-2$ in 5D MSSM,
Phys. Rev. D \textbf{94}, no.7, 075016 (2016)
doi:10.1103/PhysRevD.94.075016
[arXiv:1606.05329 [hep-ph]].
\bibitem{li46}
A.~Kobakhidze, M.~Talia and L.~Wu,
Probing the MSSM explanation of the muon g-2 anomaly in dark matter experiments and at a 100 TeV $pp$ collider,
Phys. Rev. D \textbf{95}, no.5, 055023 (2017)
doi:10.1103/PhysRevD.95.055023
[arXiv:1608.03641 [hep-ph]].
\bibitem{li47}
G.~B\'elanger, J.~Da Silva and H.~M.~Tran,
Dark matter in U(1) extensions of the MSSM with gauge kinetic mixing,
Phys. Rev. D \textbf{95}, no.11, 115017 (2017)
doi:10.1103/PhysRevD.95.115017
[arXiv:1703.03275 [hep-ph]].
\bibitem{li48}
T.~Fukuyama, N.~Okada and H.~M.~Tran,
Sparticle spectroscopy of the minimal SO(10) model,
Phys. Lett. B \textbf{767}, 295-302 (2017)
doi:10.1016/j.physletb.2017.02.021
[arXiv:1611.08341 [hep-ph]].
\bibitem{li49}
A.~Choudhury, S.~Rao and L.~Roszkowski,
Impact of LHC data on muon $g-2$ solutions in a vectorlike extension of the constrained MSSM,
Phys. Rev. D \textbf{96}, no.7, 075046 (2017)
doi:10.1103/PhysRevD.96.075046
[arXiv:1708.05675 [hep-ph]].
\bibitem{li50}
K.~Hagiwara, K.~Ma and S.~Mukhopadhyay,
Closing in on the chargino contribution to the muon g-2 in the MSSM: current LHC constraints,
Phys. Rev. D \textbf{97}, no.5, 055035 (2018)
doi:10.1103/PhysRevD.97.055035
[arXiv:1706.09313 [hep-ph]].
\bibitem{li51}
M.~Endo, K.~Hamaguchi, S.~Iwamoto and T.~Kitahara,
Muon $g-2$ vs LHC Run 2 in supersymmetric models,
JHEP \textbf{04}, 165 (2020)
doi:10.1007/JHEP04(2020)165
[arXiv:2001.11025 [hep-ph]].
\bibitem{li52}
M.~Chakraborti, S.~Heinemeyer and I.~Saha,
Improved $(g-2)_\mu$ Measurements and Supersymmetry,
Eur. Phys. J. C \textbf{80}, no.10, 984 (2020)
doi:10.1140/epjc/s10052-020-08504-8
[arXiv:2006.15157 [hep-ph]].
\bibitem{li53}
S.~I.~Horigome, T.~Katayose, S.~Matsumoto and I.~Saha,
Leptophilic fermion WIMP: Role of future lepton colliders,
Phys. Rev. D \textbf{104}, no.5, 055001 (2021)
doi:10.1103/PhysRevD.104.055001
[arXiv:2102.08645 [hep-ph]].
\bibitem{li54}
W.~Yin and N.~Yokozaki,
Splitting mass spectra and muon g \ensuremath{-} 2 in Higgs-anomaly mediation,
Phys. Lett. B \textbf{762}, 72-79 (2016)
doi:10.1016/j.physletb.2016.09.024
[arXiv:1607.05705 [hep-ph]].
\bibitem{li55}
B.~Zhu, R.~Ding and T.~Li,
Higgs mass and muon anomalous magnetic moment in the MSSM with gauge-gravity hybrid mediation,
Phys. Rev. D \textbf{96}, no.3, 035029 (2017)
doi:10.1103/PhysRevD.96.035029
[arXiv:1610.09840 [hep-ph]].
\bibitem{li56}
M.~Hussain and R.~Khalid,
Understanding the muon anomalous magnetic moment in light of a flavor symmetry-based Minimal Supersymmetric Standard Model,
PTEP \textbf{2018}, no.8, 083B06 (2018)
doi:10.1093/ptep/pty087
[arXiv:1704.04085 [hep-ph]].
\bibitem{li57}
X.~Ning and F.~Wang,
Solving the muon g-2 anomaly within the NMSSM from generalized deflected AMSB,
JHEP \textbf{08}, 089 (2017)
doi:10.1007/JHEP08(2017)089
[arXiv:1704.05079 [hep-ph]].
\bibitem{li58}
M.~Frank and \"O.~\"Ozdal,
Exploring the supersymmetric U(1)$_{B-L} \times$ U(1)$_{R}$ model with dark matter, muon $g-2$ and $Z^\prime$ mass limits,
Phys. Rev. D \textbf{97}, no.1, 015012 (2018)
doi:10.1103/PhysRevD.97.015012
[arXiv:1709.04012 [hep-ph]].
\bibitem{li59}
E.~Bagnaschi, K.~Sakurai, M.~Borsato, O.~Buchmueller, M.~Citron, J.~C.~Costa, A.~De Roeck, M.~J.~Dolan, J.~R.~Ellis and H.~Fl\"acher, \textit{et al.}
Likelihood Analysis of the pMSSM11 in Light of LHC 13-TeV Data,
Eur. Phys. J. C \textbf{78}, no.3, 256 (2018)
doi:10.1140/epjc/s10052-018-5697-0
[arXiv:1710.11091 [hep-ph]].
\bibitem{li60}
C.~Li, B.~Zhu and T.~Li,
Naturalness, dark matter, and the muon anomalous magnetic moment in supersymmetric extensions of the standard model with a pseudo-Dirac gluino,
Nucl. Phys. B \textbf{927}, 255-273 (2018)
doi:10.1016/j.nuclphysb.2017.12.012
[arXiv:1704.05584 [hep-ph]].
\bibitem{li61}
G.~Pozzo and Y.~Zhang,
Constraining resonant dark matter with combined LHC electroweakino searches,
Phys. Lett. B \textbf{789}, 582-591 (2019)
doi:10.1016/j.physletb.2018.12.062
[arXiv:1807.01476 [hep-ph]].
\bibitem{li62}
Z.~Alt\i{}n, \"O.~\"Ozdal and C.~S.~Un,
Muon g-2 in an alternative quasi-Yukawa unification with a less fine-tuned seesaw mechanism,
Phys. Rev. D \textbf{97}, no.5, 055007 (2018)
doi:10.1103/PhysRevD.97.055007
[arXiv:1703.00229 [hep-ph]].
\bibitem{li63}
M.~Chakraborti, U.~Chattopadhyay and S.~Poddar,
How light a higgsino or a wino dark matter can become in a compressed scenario of MSSM,
JHEP \textbf{09}, 064 (2017)
doi:10.1007/JHEP09(2017)064
[arXiv:1702.03954 [hep-ph]].
\bibitem{li64}
T.~T.~Yanagida and N.~Yokozaki,
Muon g \ensuremath{-} 2 in MSSM gauge mediation revisited,
Phys. Lett. B \textbf{772}, 409-414 (2017)
doi:10.1016/j.physletb.2017.07.002
[arXiv:1704.00711 [hep-ph]].
\bibitem{li65}
A.~Choudhury, L.~Darm\'e, L.~Roszkowski, E.~M.~Sessolo and S.~Trojanowski,
Muon g \ensuremath{-} 2 and related phenomenology in constrained vector-like extensions of the MSSM,
JHEP \textbf{05}, 072 (2017)
doi:10.1007/JHEP05(2017)072
[arXiv:1701.08778 [hep-ph]].
\bibitem{li66}
M.~Endo, K.~Hamaguchi, S.~Iwamoto and K.~Yanagi,
Probing minimal SUSY scenarios in the light of muon $g-2$ and dark matter,
JHEP \textbf{06}, 031 (2017)
doi:10.1007/JHEP06(2017)031
[arXiv:1704.05287 [hep-ph]].
\bibitem{li67}
K.~Wang, F.~Wang, J.~Zhu and Q.~Jie,
The semi-constrained NMSSM in light of muon g-2, LHC, and dark matter constraints,
Chin. Phys. C \textbf{42}, no.10, 103109-103109 (2018)
doi:10.1088/1674-1137/42/10/103109
[arXiv:1811.04435 [hep-ph]].
\bibitem{li68}
G.~Bhattacharyya, T.~T.~Yanagida and N.~Yokozaki,
An extended gauge mediation for muon $(g-2)$ explanation,
Phys. Lett. B \textbf{784}, 118-121 (2018)
doi:10.1016/j.physletb.2018.07.037
[arXiv:1805.01607 [hep-ph]].
\bibitem{li69}
P.~Cox, C.~Han and T.~T.~Yanagida,
Muon $g-2$ and dark matter in the minimal supersymmetric standard model,
Phys. Rev. D \textbf{98}, no.5, 055015 (2018)
doi:10.1103/PhysRevD.98.055015
[arXiv:1805.02802 [hep-ph]].
\bibitem{li70}
P.~Cox, C.~Han, T.~T.~Yanagida and N.~Yokozaki,
Gaugino mediation scenarios for muon $g-2$ and dark matter,
JHEP \textbf{08}, 097 (2019)
doi:10.1007/JHEP08(2019)097
[arXiv:1811.12699 [hep-ph]].
\bibitem{li71}
J.~L.~Yang, T.~F.~Feng, Y.~L.~Yan, W.~Li, S.~M.~Zhao and H.~B.~Zhang,
Lepton-flavor violation and two loop electroweak corrections to $(g-2)_\mu$ in the B-L symmetric SSM,
Phys. Rev. D \textbf{99}, no.1, 015002 (2019)
doi:10.1103/PhysRevD.99.015002
[arXiv:1812.03860 [hep-ph]].
\bibitem{li72}
H.~M.~Tran and H.~T.~Nguyen,
GUT-inspired MSSM in light of muon $g-2$ and LHC results at $\sqrt{s}=13$   TeV,
Phys. Rev. D \textbf{99}, no.3, 035040 (2019)
doi:10.1103/PhysRevD.99.035040
[arXiv:1812.11757 [hep-ph]].
\bibitem{li73}
F.~Wang, K.~Wang, J.~M.~Yang and J.~Zhu,
Solving the muon g-2 anomaly in CMSSM extension with non-universal gaugino masses,
JHEP \textbf{12}, 041 (2018)
doi:10.1007/JHEP12(2018)041
[arXiv:1808.10851 [hep-ph]].
\bibitem{li74}
M.~Abdughani, K.~I.~Hikasa, L.~Wu, J.~M.~Yang and J.~Zhao,
Testing electroweak SUSY for muon $g$ \ensuremath{-} 2 and dark matter at the LHC and beyond,
JHEP \textbf{11}, 095 (2019)
doi:10.1007/JHEP11(2019)095
[arXiv:1909.07792 [hep-ph]].
\bibitem{li75}
W.~Kotlarski, D.~St\"ockinger and H.~St\"ockinger-Kim,
Low-energy lepton physics in the MRSSM: $(g-2)_\mu$, $\mu \to e\gamma$ and $\mu\to e$ conversion,
JHEP \textbf{08}, 082 (2019)
doi:10.1007/JHEP08(2019)082
[arXiv:1902.06650 [hep-ph]].
\bibitem{li76}
X.~X.~Dong, S.~M.~Zhao, H.~B.~Zhang and T.~F.~Feng,
The two-loop corrections to lepton MDMs and EDMs in the EBLMSSM,
J. Phys. G \textbf{47}, no.4, 045002 (2020)
doi:10.1088/1361-6471/ab5f8f
[arXiv:1901.07701 [hep-ph]].
\bibitem{li77}
M.~Ibe, M.~Suzuki, T.~T.~Yanagida and N.~Yokozaki,
Muon $g-2$ in Split-Family SUSY in light of LHC Run II,
Eur. Phys. J. C \textbf{79}, no.8, 688 (2019)
doi:10.1140/epjc/s10052-019-7186-5
[arXiv:1903.12433 [hep-ph]].
\bibitem{li78}
J.~L.~Yang, T.~F.~Feng and H.~B.~Zhang,
Electron and muon $(g-2)$ in the B-LSSM,
J. Phys. G \textbf{47}, no.5, 055004 (2020)
doi:10.1088/1361-6471/ab7986
[arXiv:2003.09781 [hep-ph]].
\bibitem{li79}
T.~T.~Yanagida, W.~Yin and N.~Yokozaki,
Muon $g-2$ in Higgs-anomaly mediation,
JHEP \textbf{06}, 154 (2020)
doi:10.1007/JHEP06(2020)154
[arXiv:2001.02672 [hep-ph]].
\bibitem{li80}
C.~Han, M.~L.~L\'opez-Ib\'a\~nez, A.~Melis, \'O.~Vives, L.~Wu and J.~M.~Yang,
LFV and (g-2) in non-universal SUSY models with light higgsinos,
JHEP \textbf{05}, 102 (2020)
doi:10.1007/JHEP05(2020)102
[arXiv:2003.06187 [hep-ph]].
\bibitem{li81}
J.~Cao, J.~Lian, L.~Meng, Y.~Yue and P.~Zhu,
Anomalous muon magnetic moment in the inverse seesaw extended next-to-minimal supersymmetric standard model,
Phys. Rev. D \textbf{101}, no.9, 095009 (2020)
doi:10.1103/PhysRevD.101.095009
[arXiv:1912.10225 [hep-ph]].
\bibitem{li82}
M.~Yamaguchi and W.~Yin,
A novel approach to finely tuned supersymmetric standard models: The case of the non-universal Higgs mass model,
PTEP \textbf{2018}, no.2, 023B06 (2018)
doi:10.1093/ptep/pty002
[arXiv:1606.04953 [hep-ph]].
\bibitem{li83}
T.~T.~Yanagida, W.~Yin and N.~Yokozaki,
Flavor-Safe Light Squarks in Higgs-Anomaly Mediation,
JHEP \textbf{04}, 012 (2018)
doi:10.1007/JHEP04(2018)012
[arXiv:1801.05785 [hep-ph]].
\bibitem{li84}
Y.~Shimizu and W.~Yin,
Natural split mechanism for sfermions: $N =$ 2 supersymmetry in phenomenology,
Phys. Lett. B \textbf{754}, 118-124 (2016)
doi:10.1016/j.physletb.2016.01.012
[arXiv:1509.04933 [hep-ph]].
\bibitem{li85}
W.~Yin,
Fixed Point and Anomaly Mediation in Partially $N = 2$ Supersymmetric Standard Models,
Chin. Phys. C \textbf{42}, no.1, 013104 (2018)
doi:10.1088/1674-1137/42/1/013104
[arXiv:1609.03527 [hep-ph]].
\bibitem{li86}
M.~Badziak and K.~Sakurai,
Explanation of electron and muon g \ensuremath{-} 2 anomalies in the MSSM,
JHEP \textbf{10}, 024 (2019)
doi:10.1007/JHEP10(2019)024
[arXiv:1908.03607 [hep-ph]].
\bibitem{li87}
M.~Endo and W.~Yin,
Explaining electron and muon $g-2$ anomaly in SUSY without lepton-flavor mixings,
JHEP \textbf{08}, 122 (2019)
doi:10.1007/JHEP08(2019)122
[arXiv:1906.08768 [hep-ph]].
\bibitem{li88}
T.~Moroi,
The Muon anomalous magnetic dipole moment in the minimal supersymmetric standard model,
Phys. Rev. D \textbf{53}, 6565-6575 (1996)
[erratum: Phys. Rev. D \textbf{56}, 4424 (1997)]
doi:10.1103/PhysRevD.53.6565
[arXiv:hep-ph/9512396 [hep-ph]].
\bibitem{li89}
S.~M.~Zhao, T.~F.~Feng, H.~B.~Zhang, B.~Yan and X.~J.~Zhan,
The corrections from one loop and two-loop Barr-Zee type diagrams to muon MDM in BLMSSM,
JHEP \textbf{11}, 119 (2014)
doi:10.1007/JHEP11(2014)119
[arXiv:1405.7561 [hep-ph]].
\bibitem{li90}
J.~Cao, F.~Li, J.~Lian, Y.~Pan and D.~Zhang,
Impact of LHC probes of SUSY and recent measurement of $(g-2)_{\mu}$ on $\mathbb{Z}_3$-NMSSM,
[arXiv:2204.04710 [hep-ph]].

\bibitem{Cao:2022chy}
J.~Cao, J.~Lian, Y.~Pan, Y.~Yue and D.~Zhang,
Impact of recent $(g-2)_{\mu}$ measurement on the light CP-even Higgs scenario in general Next-to-Minimal Supersymmetric Standard Model,
[arXiv:2201.11490 [hep-ph]].

\bibitem{Chakraborti:2022sbj}
M.~Chakraborti, S.~Heinemeyer and I.~Saha,
SUSY Dark Matter Direct Detection Prospects based on $(g-2)_\mu$,
[arXiv:2201.03390 [hep-ph]].

\bibitem{3a3b2}
S.~K.~Vempati,
Introduction to MSSM,
[arXiv:1201.0334 [hep-ph]].

\bibitem{3a3b3}
M.~Maniatis,
The Next-to-Minimal Supersymmetric extension of the Standard Model reviewed,
Int. J. Mod. Phys. A \textbf{25} (2010), 3505-3602
[arXiv:0906.0777 [hep-ph]].

\bibitem{3a3b4}
U.~Ellwanger, C.~Hugonie and A.~M.~Teixeira,
The Next-to-Minimal Supersymmetric Standard Model,
Phys. Rept. \textbf{496} (2010), 1-77
[arXiv:0910.1785 [hep-ph]].

\bibitem{3a3b5}
J.~J.~Cao, Z.~X.~Heng, J.~M.~Yang, Y.~M.~Zhang and J.~Y.~Zhu,
A SM-like Higgs near 125 GeV in low energy SUSY: a comparative study for MSSM and NMSSM,
JHEP \textbf{03} (2012), 086
[arXiv:1202.5821 [hep-ph]].

\bibitem{42}
H.~Zhou, J.~Cao, J.~Lian and D.~Zhang,
Singlino-dominated dark matter in Z3-symmetric NMSSM,
Phys. Rev. D \textbf{104} (2021) no.1, 015017
[arXiv:2102.05309 [hep-ph]].

\bibitem{17}
J.~Cao, J.~Lian, Y.~Pan, D.~Zhang and P.~Zhu,
Improved $(g-2)_\mu$ measurement and singlino dark matter in $\mu$-term extended $\mathbb{Z}_3$-NMSSM,
JHEP \textbf{09} (2021), 175
[arXiv:2104.03284 [hep-ph]].

\bibitem{16}
J.~Cao, D.~Li, J.~Lian, Y.~Yue and H.~Zhou,
Singlino-dominated dark matter in general NMSSM,''
JHEP \textbf{06} (2021), 176
[arXiv:2102.05317 [hep-ph]].



\bibitem{29}
G.~Aad \textit{et al.} [ATLAS],
Observation of a new particle in the search for the Standard Model Higgs boson with the ATLAS detector at the LHC,
Phys. Lett. B \textbf{716} (2012), 1-29
[arXiv:1207.7214 [hep-ex]].

\bibitem{30}
S.~Chatrchyan \textit{et al.} [CMS],
Observation of a New Boson at a Mass of 125 GeV with the CMS Experiment at the LHC,
Phys. Lett. B \textbf{716} (2012), 30-61
[arXiv:1207.7235 [hep-ex]].

\bibitem{3}
G.~Aad \textit{et al.} [ATLAS and CMS],
Measurements of the Higgs boson production and decay rates and constraints on its couplings from a combined ATLAS and CMS analysis of the LHC pp collision data at $ \sqrt{s}=7 $ and 8 TeV,
JHEP \textbf{08} (2016), 045
[arXiv:1606.02266 [hep-ex]].

\bibitem{4}
G.~Aad \textit{et al.} [ATLAS],
Study of the spin and parity of the Higgs boson in diboson decays with the ATLAS detector,
Eur. Phys. J. C \textbf{75} (2015) no.10, 476
[erratum: Eur. Phys. J. C \textbf{76} (2016) no.3, 152]
[arXiv:1506.05669 [hep-ex]].

\bibitem{5}
A.~M.~Sirunyan \textit{et al.} [CMS],
Combined measurements of Higgs boson couplings in proton\textendash{}proton collisions at $\sqrt{s}=13\,\text {Te}\text {V} $,
Eur. Phys. J. C \textbf{79} (2019) no.5, 421
[arXiv:1809.10733 [hep-ex]].

\bibitem{6}
G.~Aad \textit{et al.} [ATLAS],
Combined measurements of Higgs boson production and decay using up to $80$ fb$^{-1}$ of proton-proton collision data at $\sqrt{s}=$ 13 TeV collected with the ATLAS experiment,
Phys. Rev. D \textbf{101} (2020) no.1, 012002
[arXiv:1909.02845 [hep-ex]].

\bibitem{7}
G.~Aad \textit{et al.} [ATLAS],
Test of CP invariance in vector-boson fusion production of the Higgs boson in the $\mathrm{H} \rightarrow \tau \tau$ channel in proton-proton collisions at s=13TeV with the ATLAS detector,
Phys. Lett. B \textbf{805} (2020), 135426
[arXiv:2002.05315 [hep-ex]].

\bibitem{3a3b6}
X.~F.~Han, L.~Wang and J.~M.~Yang,
Higgs pair signal enhanced in the 2HDM with two degenerate 125 GeV Higgs bosons,
Mod. Phys. Lett. A \textbf{31} (2016) no.31, 1650178
[arXiv:1509.02453 [hep-ph]].

\bibitem{3a3b7}
S.~Moretti and S.~Munir,
Two Higgs Bosons near 125 GeV in the Complex NMSSM and the LHC Run I Data,
Adv. High Energy Phys. \textbf{2015} (2015), 509847
[arXiv:1505.00545 [hep-ph]].

\bibitem{8}
S.~AbdusSalam and M.~E.~Cabrera,
Revealing mass-degenerate states in Higgs boson signals,
Eur. Phys. J. C \textbf{79} (2019) no.12, 1034
[arXiv:1905.04249 [hep-ph]].

\bibitem{34}
J.~F.~Gunion, Y.~Jiang and S.~Kraml,
Diagnosing Degenerate Higgs Bosons at 125 GeV,
Phys. Rev. Lett. \textbf{110} (2013) no.5, 051801
[arXiv:1208.1817 [hep-ph]].

\bibitem{Shang:2020uog}
L.~Shang, P.~Sun, Z.~Heng, Y.~He and B.~Yang,
``Mass-degenerate Higgs bosons near 125 GeV in the NMSSM under current experimental constraints,''
Eur. Phys. J. C \textbf{80}, no.6, 574 (2020).

\bibitem{20}
U.~Ellwanger,
Nonrenormalizable interactions from supergravity, quantum
corrections and effecive low-energy theories,
Phys. Lett. B \textbf{133} (1983), 187-191

\bibitem{21}
S.~A.~Abel,
Destabilizing divergences in the NMSSM,
Nucl. Phys. B \textbf{480} (1996), 55-72
[arXiv:hep-ph/9609323 [hep-ph]].

\bibitem{22}
C.~F.~Kolda, S.~Pokorski and N.~Polonsky,
Stabilized singlets in supergravity as a source of the mu - parameter,
Phys. Rev. Lett. \textbf{80} (1998), 5263-5266
[arXiv:hep-ph/9803310 [hep-ph]].

\bibitem{23}
C.~Panagiotakopoulos and K.~Tamvakis,
Stabilized NMSSM without domain walls,
Phys. Lett. B \textbf{446} (1999), 224-227
[arXiv:hep-ph/9809475 [hep-ph]].

\bibitem{24}
S.~Ferrara, R.~Kallosh, A.~Linde, A.~Marrani and A.~Van Proeyen,
Superconformal Symmetry, NMSSM, and Inflation,
Phys. Rev. D \textbf{83} (2011), 025008
[arXiv:1008.2942 [hep-th]].

\bibitem{25}
W.~G.~Hollik, S.~Liebler, G.~Moortgat-Pick, S.~Pa\ss{}ehr and G.~Weiglein,
Phenomenology of the inflation-inspired NMSSM at the electroweak scale,
Eur. Phys. J. C \textbf{79} (2019) no.1, 75
[arXiv:1809.07371 [hep-ph]].

\bibitem{26}
W.~G.~Hollik, C.~Li, G.~Moortgat-Pick and S.~Paasch,
Phenomenology of a Supersymmetric Model Inspired by Inflation,
Eur. Phys. J. C \textbf{81} (2021) no.2, 141
[arXiv:2004.14852 [hep-ph]].

\bibitem{37}
M.~Tanabashi \textit{et al.} [Particle Data Group],
Review of Particle Physics,
Phys. Rev. D \textbf{98} (2018) no.3, 030001

\bibitem{38}
G.~Hinshaw \textit{et al.} [WMAP],
Nine-Year Wilkinson Microwave Anisotropy Probe (WMAP) Observations: Cosmological Parameter Results,
Astrophys. J. Suppl. \textbf{208} (2013), 19
[arXiv:1212.5226 [astro-ph.CO]].

\bibitem{39}
P.~A.~R.~Ade \textit{et al.} [Planck],
Planck 2013 results. XVI. Cosmological parameters,
Astron. Astrophys. \textbf{571} (2014), A16
[arXiv:1303.5076 [astro-ph.CO]].

\bibitem{40}
A.~Achterberg, S.~Amoroso, S.~Caron, L.~Hendriks, R.~Ruiz de Austri and C.~Weniger,
A description of the Galactic Center excess in the Minimal Supersymmetric Standard Model,
JCAP \textbf{08} (2015), 006
[arXiv:1502.05703 [hep-ph]].

\bibitem{41}
S.~Profumo,
An Introduction to Particle Dark Matter.

\bibitem{12}
E.~Aprile \textit{et al.} [XENON],
Dark Matter Search Results from a One Ton-Year Exposure of XENON1T,
Phys. Rev. Lett. \textbf{121} (2018) no.11, 111302
[arXiv:1805.12562 [astro-ph.CO]].

\bibitem{12-2}
E.~Aprile \textit{et al.} [XENON],
Constraining the spin-dependent WIMP-nucleon cross sections with XENON1T,
Phys. Rev. Lett. \textbf{122}, no.14, 141301 (2019)
doi:10.1103/PhysRevLett.122.141301
[arXiv:1902.03234 [astro-ph.CO]].

\bibitem{13}
Q.~Wang \textit{et al.} [PandaX-II],
Results of dark matter search using the full PandaX-II exposure,
Chin. Phys. C \textbf{44} (2020) no.12, 125001
[arXiv:2007.15469 [astro-ph.CO]].


\bibitem{45}
M.~Badziak, M.~Olechowski and P.~Szczerbiak,
Blind spots for neutralinos in NMSSM with light singlet scalar,
PoS \textbf{PLANCK2015} (2015), 130
[arXiv:1601.00768 [hep-ph]].

\bibitem{46}
A.~Pierce, N.~R.~Shah and K.~Freese,
Neutralino Dark Matter with Light Staus,
[arXiv:1309.7351 [hep-ph]].

\bibitem{47}
M.~Badziak, M.~Olechowski and P.~Szczerbiak,
Blind spots for neutralino dark matter in the NMSSM,
JHEP \textbf{03} (2016), 179
[arXiv:1512.02472 [hep-ph]].

\bibitem{48}
M.~Badziak, M.~Olechowski and P.~Szczerbiak,
Spin-dependent constraints on blind spots for thermal singlino-higgsino dark matter with(out) light singlets,
JHEP \textbf{07} (2017), 050
[arXiv:1705.00227 [hep-ph]].

\bibitem{48a}
F.~Domingo and U.~Ellwanger,
Constraints from the Muon g-2 on the Parameter Space of the NMSSM,
JHEP \textbf{07} (2008), 079
[arXiv:0806.0733 [hep-ph]].


\bibitem{48k}
G.~Degrassi and G.~F.~Giudice,
QED logarithms in the electroweak corrections to the muon anomalous magnetic moment,
Phys. Rev. D \textbf{58} (1998), 053007
[arXiv:hep-ph/9803384 [hep-ph]].

\bibitem{49}https://easyscanhep.hepforge.org/

\bibitem{50}
F.~Staub,
SARAH,
[arXiv:0806.0538 [hep-ph]].

\bibitem{51}
F.~Staub,
SARAH 3.2: Dirac Gauginos, UFO output, and more,
Comput. Phys. Commun. \textbf{184} (2013), 1792-1809
[arXiv:1207.0906 [hep-ph]].

\bibitem{52}
F.~Staub,
SARAH 4 : A tool for (not only SUSY) model builders,
Comput. Phys. Commun. \textbf{185} (2014), 1773-1790
[arXiv:1309.7223 [hep-ph]].

\bibitem{53}
F.~Staub,
Exploring new models in all detail with SARAH,
Adv. High Energy Phys. \textbf{2015} (2015), 840780
[arXiv:1503.04200 [hep-ph]].

\bibitem{54}
W.~Porod,
SPheno, a program for calculating supersymmetric spectra, SUSY particle decays and SUSY particle production at e+ e- colliders,
Comput. Phys. Commun. \textbf{153} (2003), 275-315
[arXiv:hep-ph/0301101 [hep-ph]].

\bibitem{55}
W.~Porod and F.~Staub,
SPheno 3.1: Extensions including flavour, CP-phases and models beyond the MSSM,
Comput. Phys. Commun. \textbf{183} (2012), 2458-2469
[arXiv:1104.1573 [hep-ph]].

\bibitem{57}
P.~Bechtle, S.~Heinemeyer, O.~Stal, T.~Stefaniak and G.~Weiglein,
Applying Exclusion Likelihoods from LHC Searches to Extended Higgs Sectors,
Eur. Phys. J. C \textbf{75} (2015) no.9, 421
[arXiv:1507.06706 [hep-ph]].

\bibitem{56}
P.~Bechtle, S.~Heinemeyer, O.~St\r{a}l, T.~Stefaniak and G.~Weiglein,
Probing the Standard Model with Higgs signal rates from the Tevatron, the LHC and a future ILC,
JHEP \textbf{11} (2014), 039
[arXiv:1403.1582 [hep-ph]].

\bibitem{58}
G.~Belanger, F.~Boudjema, A.~Pukhov and A.~Semenov,
MicrOMEGAs: A Program for calculating the relic density in the MSSM,
Comput. Phys. Commun. \textbf{149} (2002), 103-120
[arXiv:hep-ph/0112278 [hep-ph]].

\bibitem{59}
G.~Belanger, F.~Boudjema, C.~Hugonie, A.~Pukhov and A.~Semenov,
Relic density of dark matter in the NMSSM,
JCAP \textbf{09} (2005), 001
[arXiv:hep-ph/0505142 [hep-ph]].

\bibitem{60}
G.~Belanger, F.~Boudjema, A.~Pukhov and A.~Semenov,
MicrOMEGAs 2.0: A Program to calculate the relic density of dark matter in a generic model,
Comput. Phys. Commun. \textbf{176} (2007), 367-382
[arXiv:hep-ph/0607059 [hep-ph]].

\bibitem{61}
G.~Belanger, F.~Boudjema, A.~Pukhov and A.~Semenov,
micrOMEGAs: A Tool for dark matter studies,
Nuovo Cim. C \textbf{033N2} (2010), 111-116
[arXiv:1005.4133 [hep-ph]].

\bibitem{62}
G.~Belanger, F.~Boudjema, A.~Pukhov and A.~Semenov,
micrOMEGAs$\_$3: A program for calculating dark matter observables,
Comput. Phys. Commun. \textbf{185} (2014), 960-985
[arXiv:1305.0237 [hep-ph]].

\bibitem{63}
D.~Barducci, G.~Belanger, J.~Bernon, F.~Boudjema, J.~Da Silva, S.~Kraml, U.~Laa and A.~Pukhov,
Collider limits on new physics within micrOMEGAs$\_$4.3,
Comput. Phys. Commun. \textbf{222} (2018), 327-338
[arXiv:1606.03834 [hep-ph]].

\bibitem{68}
C.~K.~Khosa, S.~Kraml, A.~Lessa, P.~Neuhuber and W.~Waltenberger,
SModelS Database Update v1.2.3,
LHEP \textbf{2020} (2020), 158
[arXiv:2005.00555 [hep-ph]].

\bibitem{69}
S.~Kraml, S.~Kulkarni, U.~Laa, A.~Lessa, W.~Magerl, D.~Proschofsky-Spindler and W.~Waltenberger,
SModelS: a tool for interpreting simplified-model results from the LHC and its application to supersymmetry,
Eur. Phys. J. C \textbf{74} (2014), 2868
[arXiv:1312.4175 [hep-ph]].

\bibitem{70}
F.~Ambrogi, S.~Kraml, S.~Kulkarni, U.~Laa, A.~Lessa, V.~Magerl, J.~Sonneveld, M.~Traub and W.~Waltenberger,
SModelS v1.1 user manual: Improving simplified model constraints with efficiency maps,
Comput. Phys. Commun. \textbf{227} (2018), 72-98
[arXiv:1701.06586 [hep-ph]].

\bibitem{71}
J.~Dutta, S.~Kraml, A.~Lessa and W.~Waltenberger,
SModelS extension with the CMS supersymmetry search results from Run 2,
LHEP \textbf{1} (2018) no.1, 5-12
[arXiv:1803.02204 [hep-ph]].

\bibitem{72}
J.~Heisig, S.~Kraml and A.~Lessa,
Constraining new physics with searches for long-lived particles: Implementation into SModelS,
Phys. Lett. B \textbf{788} (2019), 87-95
[arXiv:1808.05229 [hep-ph]].

\bibitem{73}
G.~Alguero, S.~Kraml and W.~Waltenberger,
A SModelS interface for pyhf likelihoods,
Comput. Phys. Commun. \textbf{264} (2021), 107909
[arXiv:2009.01809 [hep-ph]].

\bibitem{64}
W.~Beenakker, R.~Hopker and M.~Spira,
PROSPINO: A Program for the production of supersymmetric particles in next-to-leading order QCD,
[arXiv:hep-ph/9611232 [hep-ph]].

\bibitem{65}
W.~Beenakker, M.~Kramer, T.~Plehn, M.~Spira and P.~M.~Zerwas,
Stop production at hadron colliders,
Nucl. Phys. B \textbf{515} (1998), 3-14
[arXiv:hep-ph/9710451 [hep-ph]].

\bibitem{66}
W.~Beenakker, M.~Klasen, M.~Kramer, T.~Plehn, M.~Spira and P.~M.~Zerwas,
The Production of charginos / neutralinos and sleptons at hadron colliders,
Phys. Rev. Lett. \textbf{83} (1999), 3780-3783
[erratum: Phys. Rev. Lett. \textbf{100} (2008), 029901]
[arXiv:hep-ph/9906298 [hep-ph]].
	
\bibitem{67}
W.~Beenakker, R.~Hopker, M.~Spira and P.~M.~Zerwas,
Squark and gluino production at hadron colliders,
Nucl. Phys. B \textbf{492} (1997), 51-103
[arXiv:hep-ph/9610490 [hep-ph]].

\bibitem{pdg-2020}
P.~A.~Zyla \textit{et al.} [Particle Data Group],
Review of Particle Physics,
PTEP \textbf{2020} (2020) no.8, 083C01.

\end{small}
\end{thebibliography}

\end{document}